\documentclass[preprint,12pt]{elsarticle}

\usepackage{graphicx}
\usepackage{amssymb}
\usepackage{amsthm}
\usepackage{hyperref}
\usepackage{multirow}
\usepackage{color}

\newcommand{\eqref}[1]{(\ref{eq:#1})}

\newcommand{\bmath}[1]{\mbox{\boldmath$#1$}}
\newcommand{\vek}[1]{\bmath{#1}}

\newcommand{\trn}{{\sf ^T}}

\newcommand{\avgs}[1]{\left\langle #1 \right\rangle} 

\newcommand{\tenss}[1]{\bmath{#1}}                   
\newcommand{\tensf}[1]{\bmath{\mathsf{#1}}}          

\newcommand{\newtext}[1]{{\color{red}{#1}}}

\journal{Applied Mathematics and Computation}

\begin{document}

\begin{frontmatter}

\title{Homogenization of coupled heat and moisture transport in masonry structures including interfaces}

\author[ctu,cideas]{Jan S\'{y}kora}
\ead{jan.sykora.1@fsv.cvut.cz}
\author[ctu,cideas]{Michal \v{S}ejnoha\corref{auth}}
\ead{sejnom@fsv.cvut.cz}
\author[cideas]{Ji\v{r}\'{\i} \v{S}ejnoha}
\ead{sejnoha@fsv.cvut.cz}
\cortext[auth]{Corresponding author. Tel.:~+420-2-2435-4494;fax~+420-2-2431-0775}
\address[ctu]{Department of Mechanics, Faculty of Civil Engineering,
  Czech Technical University in Prague, Th\' akurova 7, 166 29 Prague
  6, Czech Republic}
\address[cideas]{Centre for Integrated Design of Advances Structures,
  Th\' akurova 7, 166 29 Prague 6, Czech Republic}

\begin{abstract}
Homogenization of a simultaneous heat and moisture flow in a
masonry wall is presented in this paper. The principle objective
is to examine an impact of the assumed imperfect hydraulic contact
on the resulting homogenized properties. Such a contact is
characterized by a certain mismatching resistance allowing us to
represent a discontinuous evolution of temperature and moisture
fields across the interface, which is in general attributed to
discontinuous capillary pressures caused by different pore size
distributions of the adjacent porous materials. In achieving this,
two particular laboratory experiments were performed to provide
distributions of temperature and relative humidity in a sample of
the masonry wall, which in turn served to extract the
corresponding jumps and subsequently to obtain the required
interface transition parameters by matching numerical predictions
and experimental results. The results suggest a low importance of
accounting for imperfect hydraulic contact for the derivation of
macroscopic homogenized properties. On the other hand, they
strongly support the need for a fully coupled multi-scale analysis
due to significant dependence of the homogenized properties on
actual moisture gradients and corresponding values of both
macroscopic temperature and relative humidity.
\end{abstract}


\end{frontmatter}

\section{Introduction}
\label{sec:intro} An extensive three-dimensional nonlinear
thermo-mechanical analysis of Charles Bridge in Prague, as a
typical representative of historical masonry structures,
identified the year round variation of temperature as one of the
most severe contributors to the growth of damage in the
bridge~\cite{Novak:ES:2007}. In sum, the analysis was carried out
in the framework of totally uncoupled multi-scale solution
strategy assuming a macroscopically homogeneous bridge with
material data derived from an independent homogenization study
performed on meso-scale. When limiting attention to a heat
transport phenomenon such a simplification has been supported by
calculations presented in~\cite{Sejnoha:MS:2008} rendering the
macroscopic homogenized heat conductivities independent of
macroscopic gradients. On the other hand, the obtained results
have shown a relatively strong dependence of the macroscopic
thermal conductivity on initial values of relative humidity and
temperature.

It will be demonstrated that in view of this the uncoupled multi-scale
approach is no longer admissible and the bridging of scales must be
understood in a fully coupled framework. Such an analysis thus
accommodates two sources of coupling on both material and structural
level. \newtext{The latter one is typically presented in the framework
  of FE$^2$ computational
  scheme~\cite{Ozdemir:IJNME:2008,Ozdemir:CMAME:2008,Larsson:IJNME:2010,Larsson:IJNAMG:2010,Lee:CM:2009}.
  Nevertheless, if coupling on a material level is considered it
  usually comprises mechanical response and transport of one
  particular non-mechanical field, either temperature or moisture.  On
  the other hand, coupling heat and moisture transport within FE$^2$
  scheme appears novel. This topic, however, goes beyond the present
  scope and will be considered elsewhere.}  Instead, our attention
will concentrate on the following two aspects of the modeling of
masonry structures:
\begin{itemize}
\item To assess the influence of possibly imperfect hydraulic contact
  on macroscopic homogenized properties. The notion of imperfect
  hydraulic contact has been put forward in~\cite[to cite a
    few]{Freitas:BE:1996,Qiu:JTEBS:2003} offering, through
  experimental observations, the discontinuity in moisture field to be
  caused by discontinuity in capillary pressure at the interface of
  distinct porous materials in the case when the corresponding pore
  size distributions do not sufficiently interpenetrate. Experimental
  validation of this assumption is presented in Section~\ref{sec:IP}.
\item To confirm the need for a fully coupled multi-scale analysis.
\end{itemize}

A crucial point in addressing the first item is the choice of a
suitable constitutive model. Literature offers a variety of such
models allowing for modeling of heat and moisture transport in
porous materials. An extensive overview of various models is
available in~\cite{Cerny:2002}. Since moisture transport due to
gravity forces can usually be neglected in building structures
including bridges we lump one particular diffusion model proposed
by K\"{u}nzel~\cite{Kunzel:R:1995}, see
also~\cite{Kunzel:IJHMT:1997} for additional reference. Because of
lucidity, a short overview of this model is provided in
Section~\ref{sec:LCE}. The model is utilized in
Section~\ref{sec:IP} to reproduce the transient transport of heat
and moisture induced in the laboratory within a sample of masonry
wall for specific climatic conditions. The desired interface
transition parameters are estimated in parallel by matching
numerical results and experimental measurements.

The second item moves our attention to the topic of homogenization. In
general, the macro-meso transition of transient heat flow calls for
the solution of the transient problem on both scales. As suggested
in~\cite{Larsson:IJNME:2010,Larsson:IJNAMG:2010}, this becomes
particularly important for a finite size representation of a sub-scale
problem (RVE - representative volume element) since estimates of
distributions of macroscopic fields may show size dependency due to
higher order terms appearing on the left hand side of macroscopic
balance equations. The authors further showed that a steady state
solution is recovered for infinitesimally small RVEs. This was an a
priori assumption put forward, e.g.
in~\cite{Ozdemir:IJNME:2008}. Because the problem of macro-scale
transient flow is not part of this contribution, the steady state
conditions will also be adopted here to see, through a parametric
study performed solely on the meso-scale, a significant dependence of
homogenized properties on applied moisture gradients thus supporting
the advocated need for a fully coupled multi-scale analysis in the
solution of real engineering problems. In doing so, we limit our
attention to classical first order homogenization theory as presented
for example
in~\cite{Sejnoha:MS:2008,Michel:1999:EPC,Sejnoha:SEM:2008}.
Individual steps are outlined in Section~\ref{sec:HMS}. The most
essential results are then summarized in Section~\ref{sec:conclusion}.

In the following text, $\vek{a}$ and $\tensf{A}$ denote a vector and a
symmetric second-order tensor, respectively. The symbol
$\vek{\nabla}=\left\{\displaystyle{{\partial}/{\partial{x}}},
\displaystyle{{\partial}/{\partial{y}}},\displaystyle{{\partial}/{\partial{z}}}\right\}^{\sf
  T}$ stands for the gradient representation. All materials are
assumed locally isotropic.

\section{Local constitutive and balance equations}
\label{sec:LCE}
Owing to its ability to describe all substantial phenomena of heat and
moisture transport in building materials with numerical predictions
reasonably close to experimentally obtained data we choose the model
developed by K\"{u}nzel~\cite{Kunzel:R:1995} for studying these
transport processes in masonry structures.

He neglected the liquid water and water vapor convection driven by
gravity and total pressure as well as enthalpy changes due to liquid
flow and choose relative humidity $\varphi$ as the only moisture
potential. The water vapor diffusion is then described by Fick's law
written as
\begin{equation}
\vek{g}_{v}=-\delta_{p}\vek{\nabla}\left(\varphi p_{\mathrm{sat}}\right),\label{eq:TFlux-1}
\end{equation}
where $\vek{g}_v$ is the water vapor flux, $\delta_{p}$ is the water
vapor permeability of a porous material and
$p_{\mathrm{sat}}=p_{\mathrm{sat}}(\theta)$ is the saturation water
vapor pressure being exponentially dependent on temperature. The
transport of liquid water is assumed in the form of surface diffusion
in an absorbed layer and capillary flow typically represented by
Kelvin's law
\begin{equation}
\vek{g}_{w} = -D_{\varphi}\vek{\nabla}\varphi,\label{eq:TFlux-2}
\end{equation}
where $\vek{g}_w$ is the flux of liquid water,
$D_{\varphi}=D_{w}\,(\mathrm{d}w/\mathrm{d}\varphi)$ is the liquid
conductivity, $D_{w}=D_{w}\,(w/w_{f})$ is the liquid diffusivity,
$\mathrm{d}w/\mathrm{d}\varphi$ is the derivative of water retention
function and $w/w_{f}$ is the water content related to the capillary
saturation with $w_{f}$ being the free water saturation. The Fourier
law is then used to express the heat flux $\vek{q}$ as
\begin{equation}
\vek{q} = -\lambda\vek{\nabla}\theta,\label{eq:TFlux-3}
\end{equation}
where $\lambda$ is the thermal conductivity and $\theta$ is the local
temperature. Introducing the above constitutive equations into energy
and mass conservation equations we finally get

\begin{itemize}
\item The energy balance equation
\begin{equation}
\frac{\mathrm{d}H}{\mathrm{d}\theta}\frac{\mathrm{d}\theta}{\mathrm{d}t}
 =  \vek{\nabla}\trn[\lambda\vek{\nabla}\theta]+h_{v}\vek{\nabla}\trn
[\delta_{p}\vek{\nabla}\left(\varphi p_{\mathrm{sat}}\right)],
\label{eq:TE3}
\end{equation}
\item The conservation of mass equation
\begin{equation}
\frac{\mathrm{d}w}{\mathrm{d}\varphi}\frac{\mathrm{d}
\varphi}{\mathrm{d}t}  =  \vek{\nabla}\trn[D_{\varphi}\vek{\nabla}\varphi]+\vek{\nabla}\trn
[\delta_{p}\vek{\nabla}\left(\varphi
p_{\mathrm{sat}}\right)],\label{eq:TE4}
\end{equation}
\end{itemize}
where $H$ is the enthalpy of moist building material and $h_{v}$ is
the evaporation enthalpy of water. The second term on the right
hand side of Eq.~\eqref{TE3} represents the change of enthalpy due
to phase transition being considered the only heat source or sink.

\section{Evaluation of interface transition parameters}
\label{sec:IP}
As foreshadowed in the introductory part, one of the objectives of the
present contribution is to evaluate the influence of imperfect
hydraulic contact on the predictions of macroscopic homogenized
transport parameters. Unlike classical definition of a hydraulic
contact, which builds upon continuity of capillary pressure $p_c$
resulting in turn into a jump of water content $w$ across the
interface uniquely related to $p_c=p_c(w)$, an imperfect contact
allows for a discontinuous variation of capillary pressure along the
interface caused by different pore size distributions of the adjacent
porous materials~\cite{Qiu:JTEBS:2003}. In case of natural contact,
assumed henceforth for the brick-mortar interface, the flux of water
vapor is neglected and the flux of liquid water $g_{w,\mathrm{int}}$
becomes
\begin{equation}
g_{w,\mathrm{int}}=-\beta_{\mathrm{int}}(p_{c2}-p_{c1}),\label{eq:jump_phi}
\end{equation}
where $\beta_{\mathrm{int}}$ is the internal interface
permeability. \newtext{The pore size difference is implicitly
  introduced through the Kelvin-Laplace equation~\cite{Kunzel:R:1995}
  yielding the capillary pressure as a function of relative humidity
  as}
\begin{equation}
p_c=-\frac{\rho^wR\theta}{M_w}{\rm ln}\varphi,\label{eq:jump_Kelvin}
\end{equation}
where $M_w$ is the molar mass of water, $\rho^w$ is the water
intrinsic density and $R$ is the universal gas constant. Although
temperature continuity is typically assumed for
natural~\cite{Freitas:BE:1996}, we shall expect that the jump in
capillary pressure results in the corresponding jump in
temperature. The heat flux across the interface $q_{\mathrm{int}}$
will then attain the form similar to Eq.~\eqref{jump_phi}
\begin{equation}
q_{\mathrm{int}}=-\alpha_{\mathrm{int}}(\theta_{2}-\theta_{1}),\label{eq:jump_theta}
\end{equation}
where $\alpha_{\mathrm{int}}$ is the internal heat transfer
coefficient.
In numerical calculations Eqs.~\eqref{jump_phi} and~\eqref{jump_theta}
are introduced by employing standard interface elements. 

\subsection{Experimental measurements}\label{sec:EO}
An extensive experimental investigation of the moisture and heat
transport through a sample of the masonry wall was carried out to
determine the previously introduced interface transport
coefficients. \newtext{Although their direct measurement is currently
  not feasible, their estimates are still possible at least in the
  context of inverse methods combining available experimental data and
  numerical simulations of corresponding laboratory tests. This
  particular approach, where model parameters are properly adjusted
  during repeated calculations to match experimental and numerical
  results, is adopted thereinafter.}

\begin{figure} [ht!]
\begin{center}
\begin{tabular}{c@{\hspace{10mm}}c}
\includegraphics*[width=40mm,keepaspectratio]{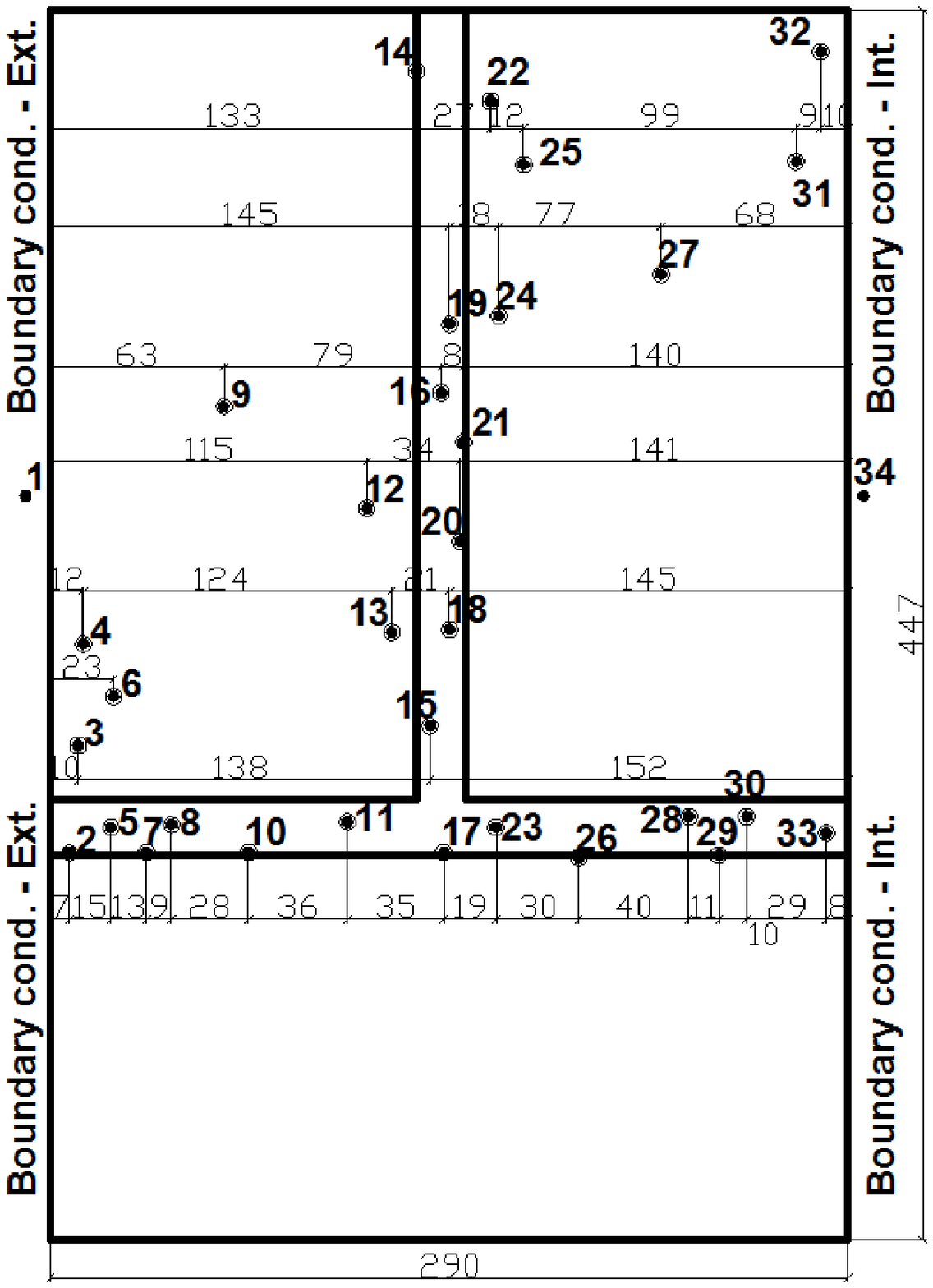}&
\includegraphics*[width=50mm,keepaspectratio]{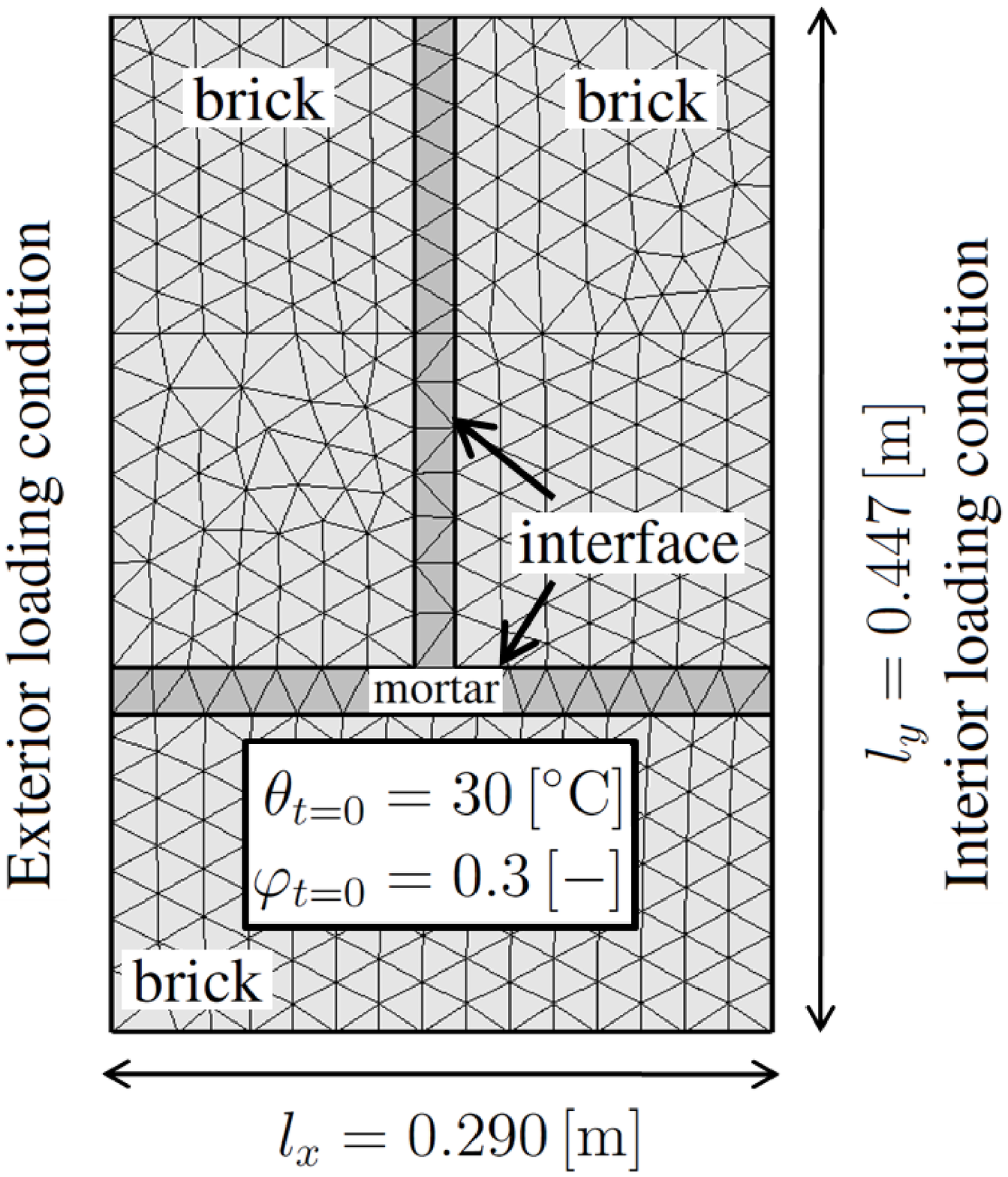}\\
(a)&(b)
\end{tabular}
\end{center}
\caption{(a) Geometry of the masonry block with positions of sensors,
  (b) finite element mesh with applied loading}
\label{fig:EO1}
\end{figure}

The experiments were conducted using the NONSTAT measuring system
consisting of two climatic chambers for the simulation of climatic
conditions (relative humidity, temperature), which are connected by a
specially developed tunnel for testing large specimens. Although the
specimen dimensions are therefore close to a real masonry structure,
the accuracy of measurements remains the same as for small laboratory
samples, see~\cite{Pavlik:JTEBS:2002,Pavlik:SFR:2010} for further
details. To measure temperature and moisture fields, a set of sensors
was attached to the masonry specimen as seen in Fig.~\ref{fig:EO1}(a)
with the following accuracy: capacitive relative humidity sensors are
applicable in the range of humidities $0.05$-$0.98$ $\pm$ $0.02$ [-],
temperature sensors provide measurements with a deviation of $\pm$
$0.4$ [$^{\circ}\mathrm{C}$] in the temperature range from -20 to 0
[$^{\circ}\mathrm{C}$] and $\pm$ $0.1$ [$^{\circ}\mathrm{C}$] in the
range from $0$ to $70$ [$^{\circ}\mathrm{C}$].

\begin{figure}[ht]
\begin{center}
\begin{tabular}{cc}
\includegraphics*[width=65mm,keepaspectratio]{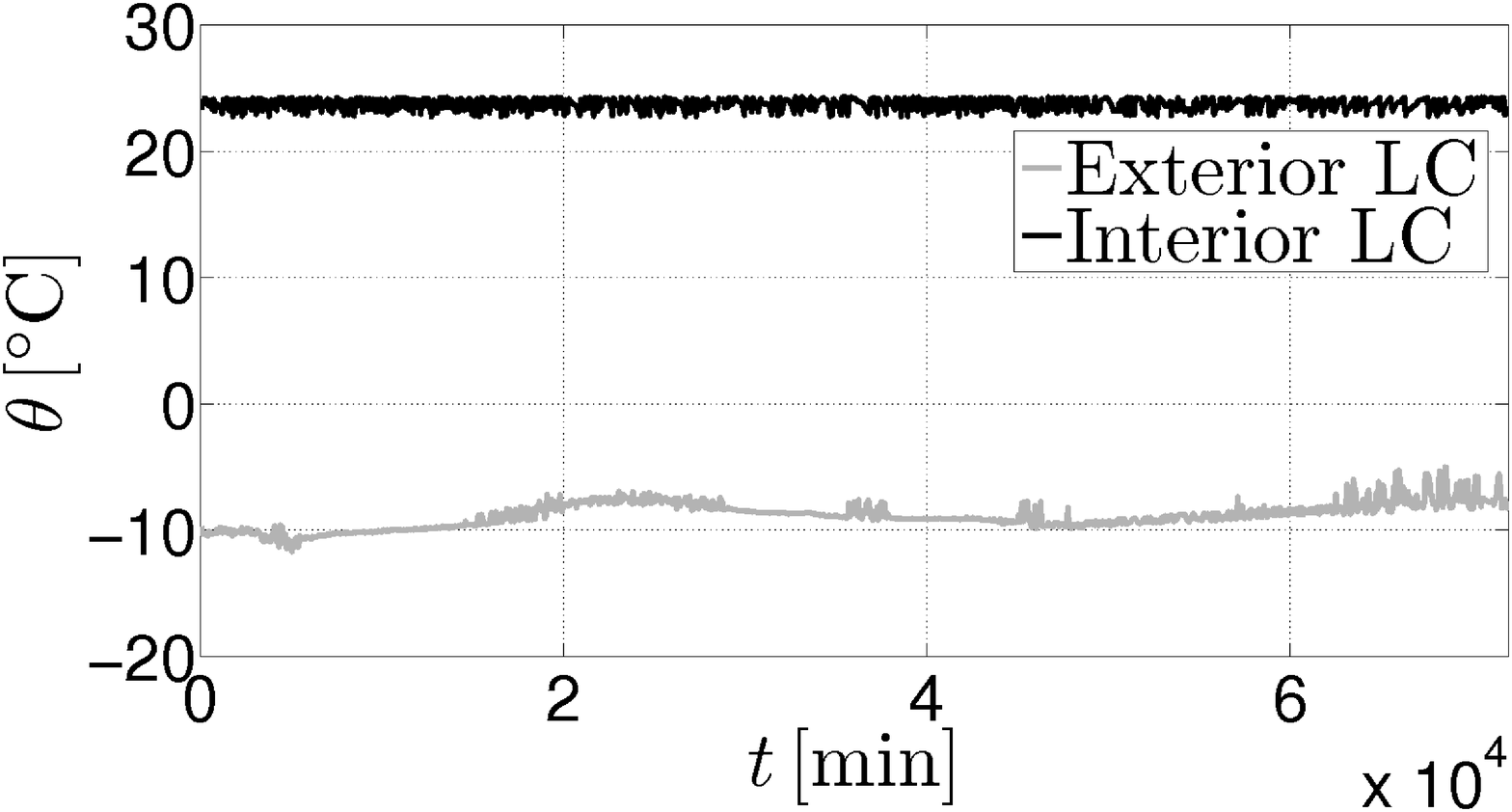}&
\includegraphics*[width=65mm,keepaspectratio]{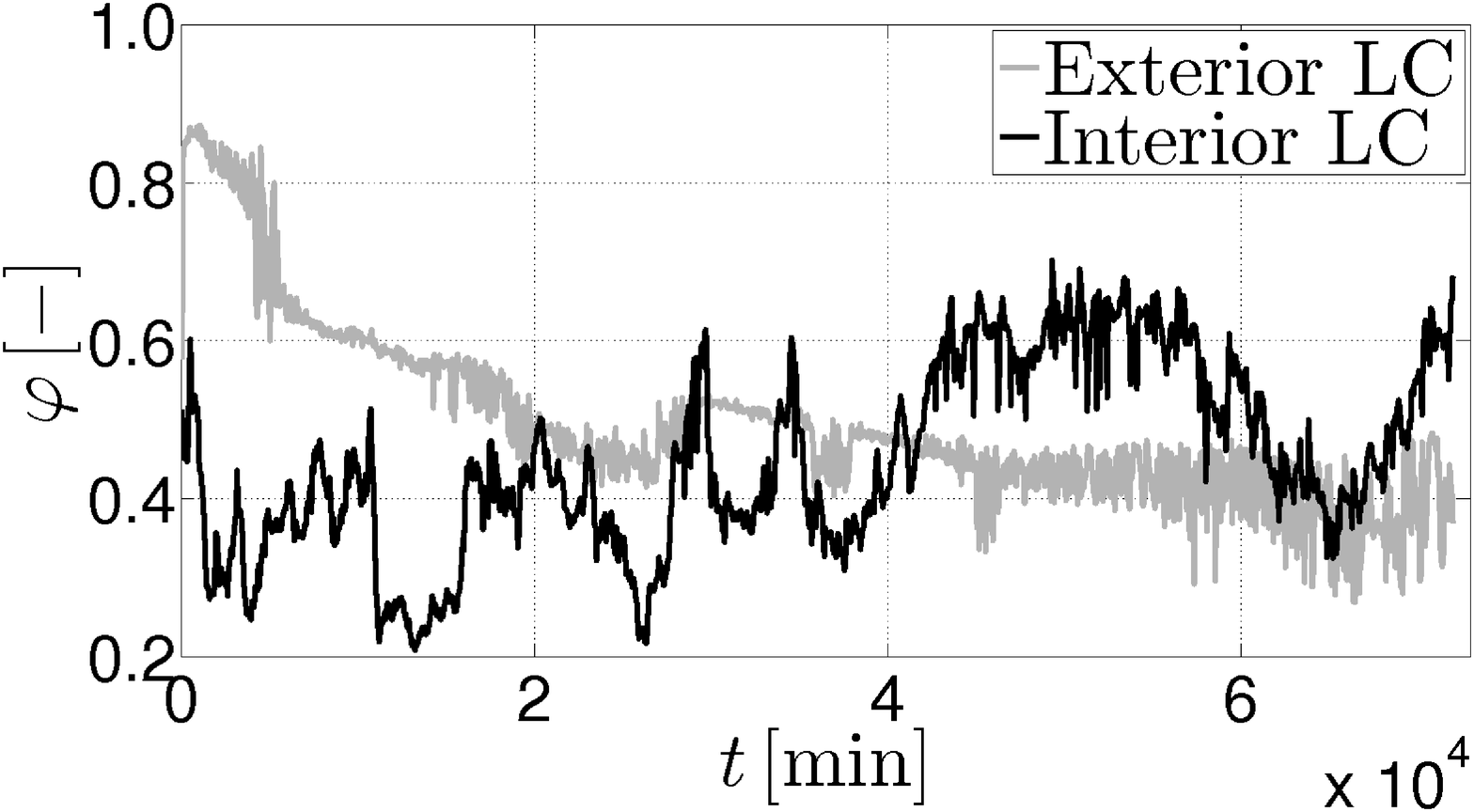}\\
(a) & (b)
\end{tabular}
\caption{Experiment No. 1:
  Loading conditions - (a) temperature, (b) relative
  humidity} \label{fig:NC1}
\end{center}
\begin{center}
\begin{tabular}{cc}
\includegraphics*[width=65mm,keepaspectratio]{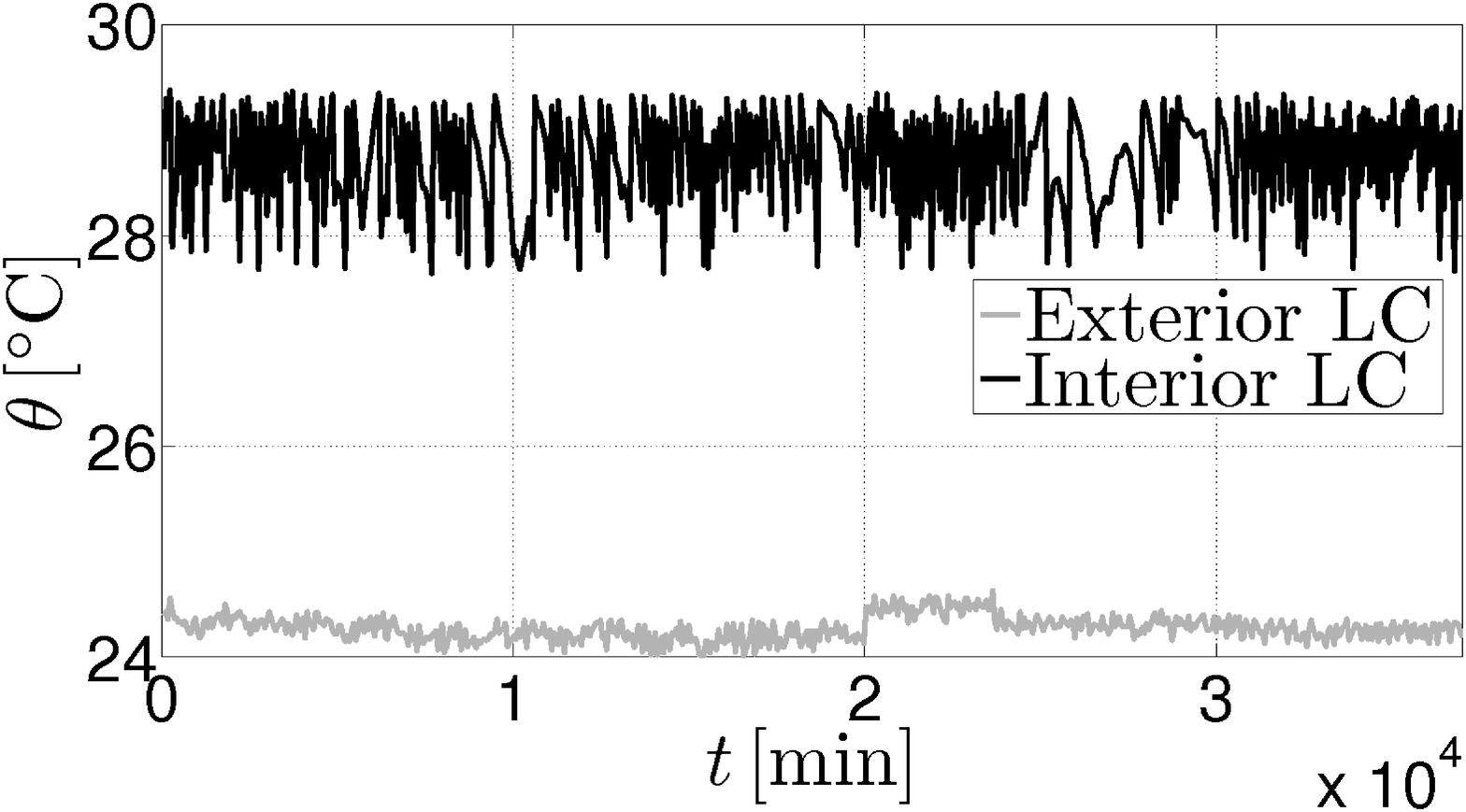}
& \includegraphics*[width=65mm,keepaspectratio]{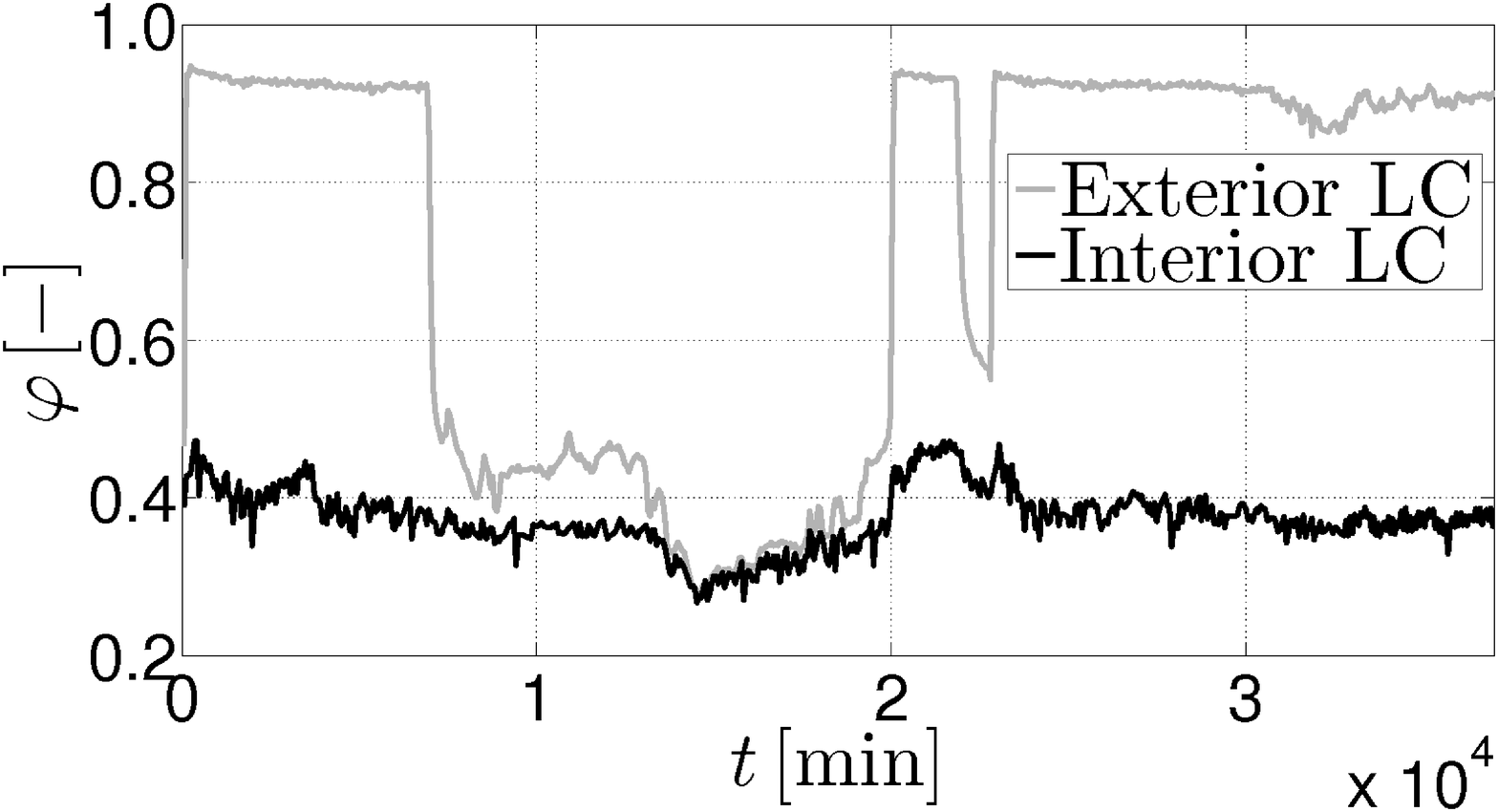}\\
(a) & (b)\\
\end{tabular}
\caption{Experiment No. 2: Loading conditions - (a) temperature, (b)
  relative humidity} \label{fig:NC5}
\end{center}
\end{figure}

Two separate experiments were conducted. In the first experiment, the
heat transport was driven by the temperature gradient, whereas the
relative humidity was maintained at a constant level. In the second
experiment, the relative humidity transport was monitored at
approximately constant temperature. The real climatic conditions
generated on both the interior and exterior part of the wall sample
for the first and second experiment are displayed in
Figs.~\ref{fig:NC1} and~\ref{fig:NC5}, respectively. \newtext{Notice
  that although expected to be constant around the value of 0.5, the
  relative humidity in the first experiment shows considerable
  fluctuations exceeding the range of 0.2 [-]. Moreover, the
  measurements in the second experiments were polluted by two
  electricity shutdowns clearly visible in Fig.~\ref{fig:NC5}
  resulting in a severe drop of relative humidity on the exterior part
  of the specimen. It should, however, be mentioned that despite of
  that, none of these two complications in the control of relative
  humidity are of major concern in the identification of material
  parameters since the exact record of boundary conditions, with
  limitations to the selected time step, was introduced in all
  numerical calculations.}

\begin{figure}
\begin{center}
\begin{tabular}{cc}
\includegraphics*[width=65mm,keepaspectratio]{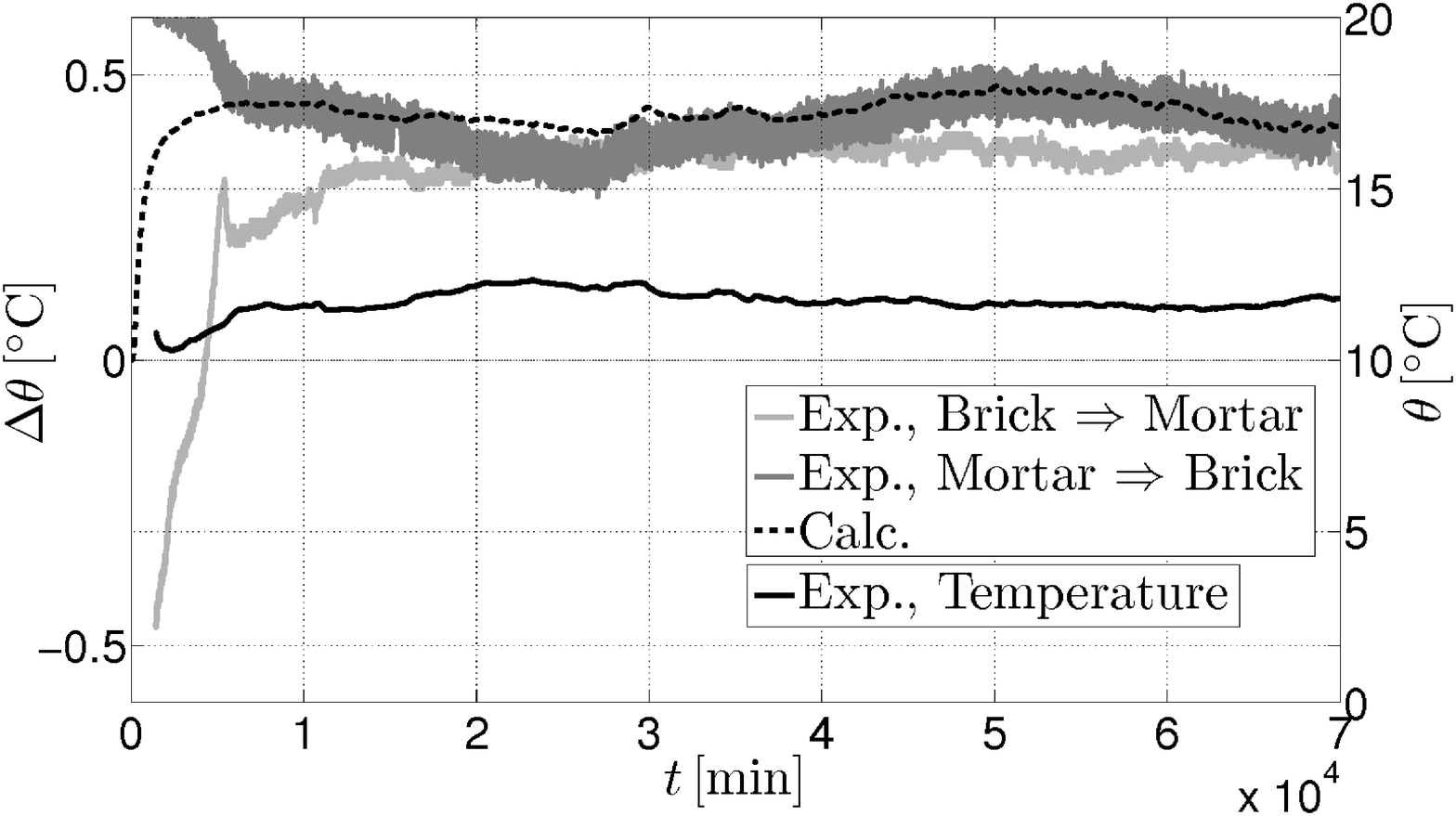}&
\includegraphics*[width=65mm,keepaspectratio]{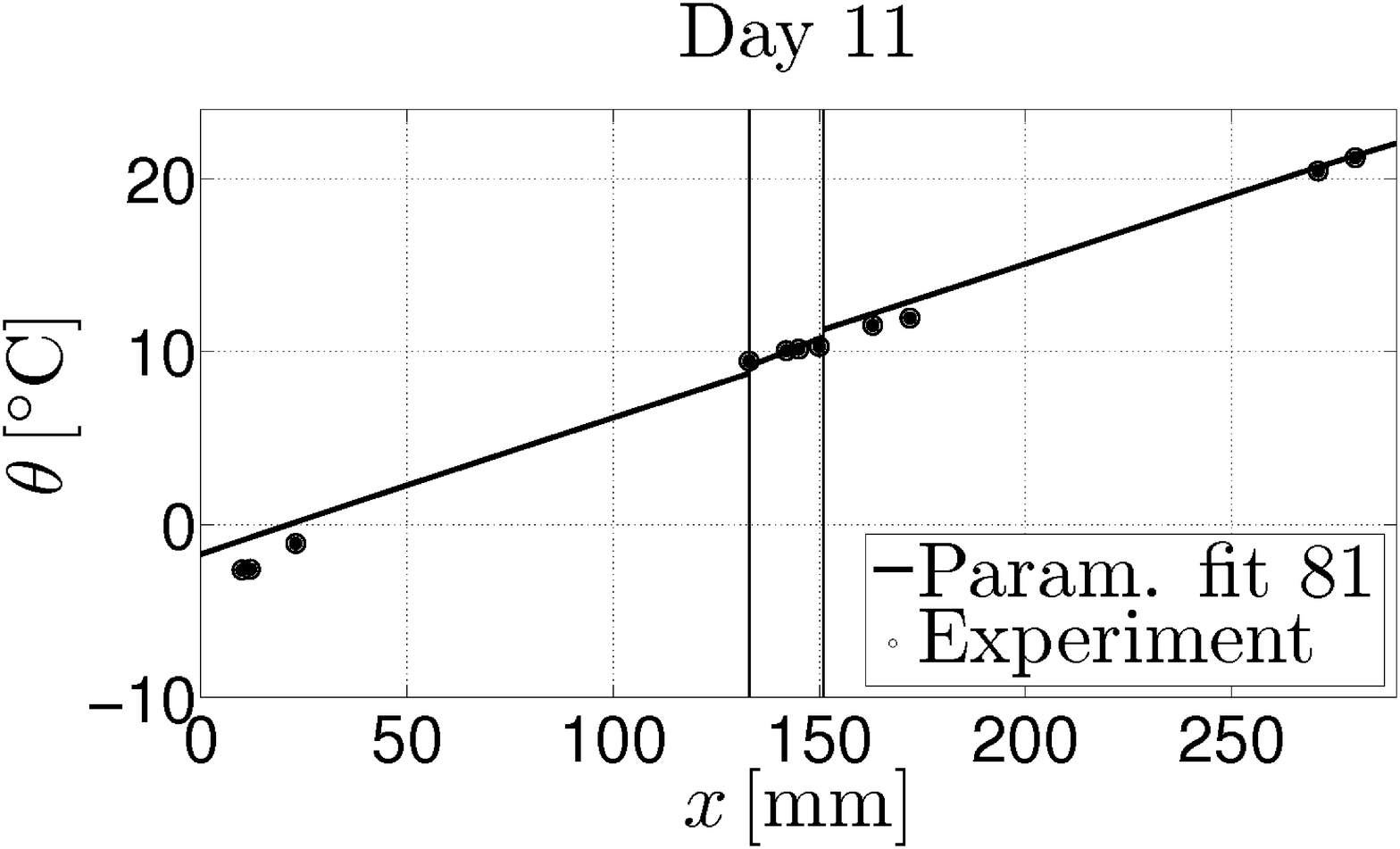}\\
(a)&(b)
\end{tabular}
\caption{Experiment No. 1: (a) comparison of experimentally and
  numerically obtained temperature jumps along the interface, (b)
  calculated and measured temperature profile in chosen sensors
  corresponding to $11^\mathrm{th}$ day}
\label{fig:NC2}
\end{center}
\end{figure}

\subsection{Numerical simulation}
The first experiment was focused on the heat transport trough the
masonry block, especially in the area of interface transition zone. On
the exterior side of the specimen, a constant temperature of $-9.5$
[$^{\circ}\mathrm{C}$] and on the interior side, a constant
temperature of $24.5$ [$^{\circ}\mathrm{C}$] were maintained, see
Fig.~\ref{fig:NC1}(a). The experimental measurement lasted $50$ days.
\newtext{All experiments were performed considering the moisture in
  the form of water vapor only. Therefore, no additional difficulties
  associated with ice formation in large pores aroused in numerical
  simulations. Nevertheless, if this issue becomes important the
  present model can be modified, as suggested in~\cite{Kunzel:R:1995},
  allowing us to determine the amount of movable water for a given
  temperature.}

Distribution of temperature in the selected points close to the
interface served to extract the corresponding jumps and to assess
their dependence on temperature and relative humidity on the one hand
and on the other hand their sensitivity to the direction of flow.
\newtext{The experimentally obtained time variation of temperature
  jumps is plotted in Fig.~\ref{fig:NC2} suggesting its invariance
  with respect to both temperature and flow direction, at least for
  the present experimental setup.} An example of variation of
temperature is plotted in Fig.~\ref{fig:NC2}(a) for sensor
No. 12. Reproducing these results numerically would thus allow us to
derive the respective interface heat transfer coefficient
$\alpha_{\mathrm{int}}$.

\begin{figure} [ht]
\begin{center}
\begin{tabular}{cc}
\includegraphics*[width=65mm,keepaspectratio]{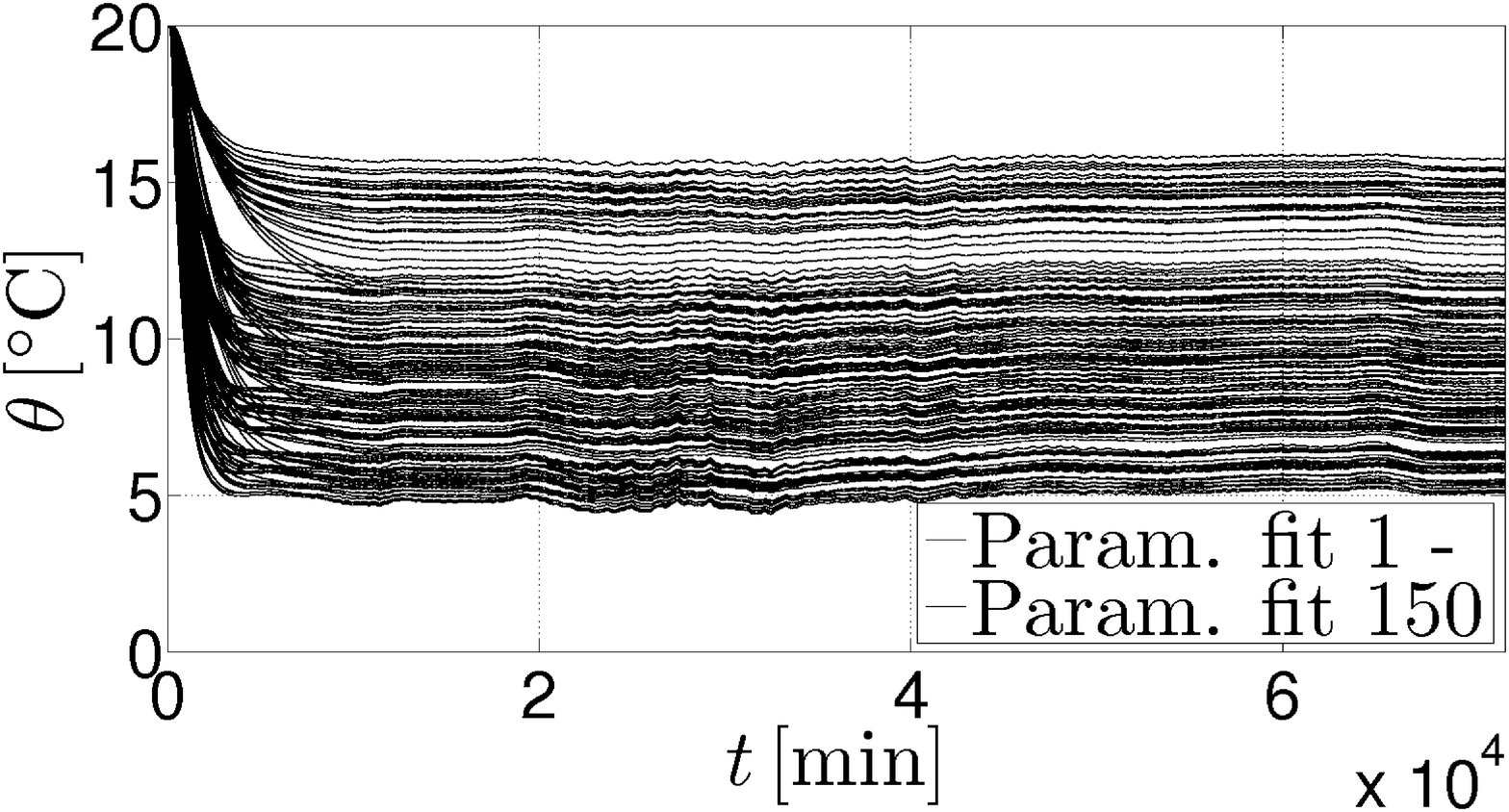}
& \includegraphics*[width=65mm,keepaspectratio]{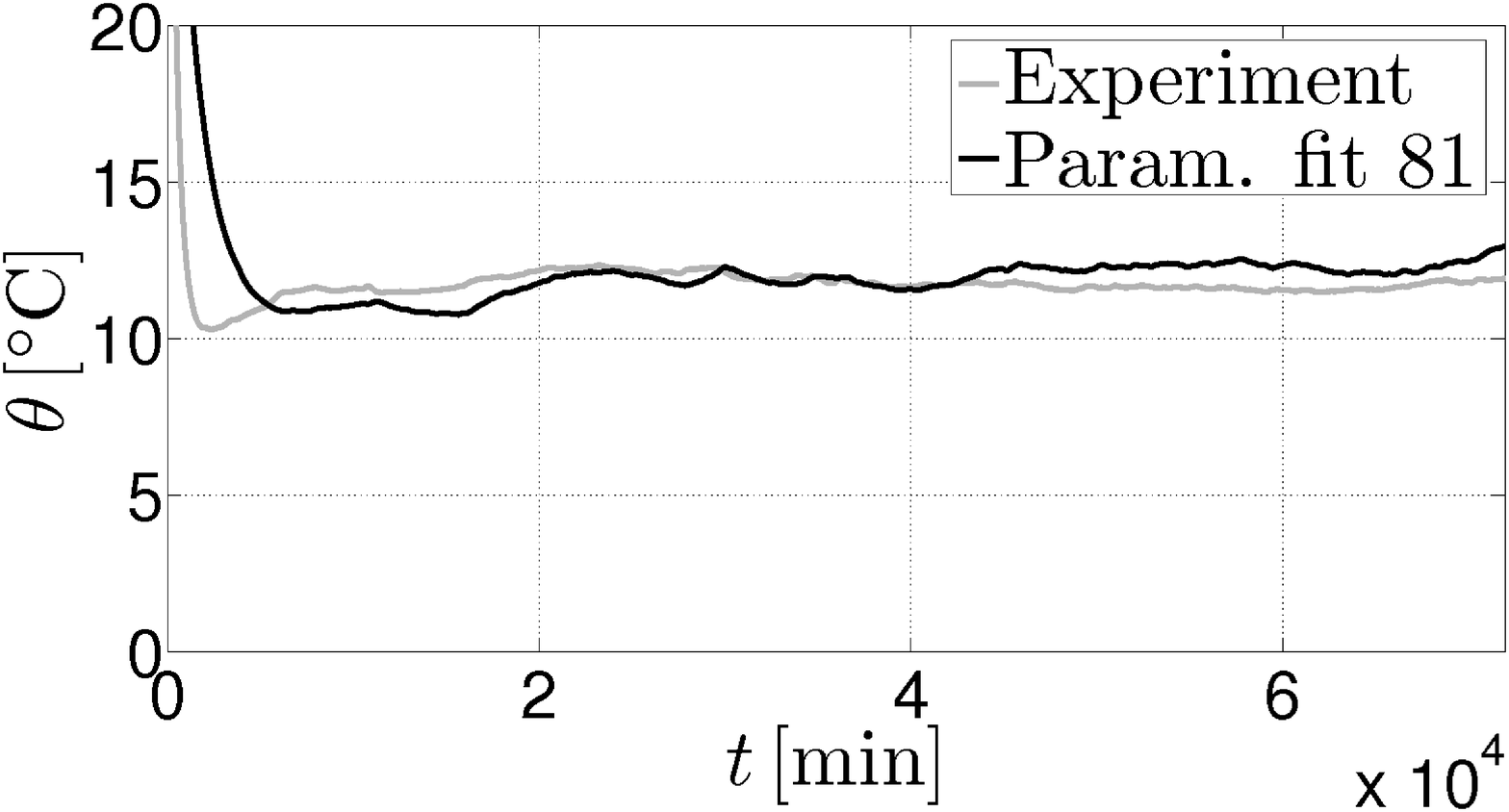}\\
(a) & (b)
\end{tabular}
\caption{Experiment No. 1: (a) evolution of temperature obtained from
  all numerical calculations in sensor No. $12$, (b) evolution of
  temperature obtained numerically for optimal data and from
  experiment in sensor No. $12$} \label{fig:NC4}
\end{center}
\end{figure}

If material parameters of individual phases were available and
assuming the relative humidity is continuous this would be the only
parameter to be searched for. Unfortunately, no additional experiments
were conducted for this type of material and had to be identified
along with the parameter $\alpha_{\mathrm{int}}$.  Although a variety
of techniques, typically exploiting the power of genetic algorithms,
are available in the literature, see e.g.
~\cite{Matous:2000:GEI,Kucerova:2007:PHD}, a simple "trial and error"
procedure was adopted here to mine the necessary material data. To
that end, the moisture transport parameters were taken from literature
for a similar type of material~\cite{Sejnoha:MS:2008}, while the
remaining thermal properties of brick, mortar and interface transition
zone, namely the thermal conductivity of dry building material
$\lambda_{\mathrm{0}}$ \newtext{[Wm$^{-1}$K$^{-1}$]}, the thermal
conductivity supplement $b_{\mathrm{tcs}}$ \newtext{[-]} and the
interface heat transfer coefficient $\alpha_{\mathrm{int}}$
\newtext{[Wm$^{-2}$K$^{-1}$]} were found by matching the experimental
and numerical results in the framework of least square method applied
to a pool of numerical realizations with input parameters generated by
the Latin Hypercube Sampling method assuming Log-normal distribution
of material data \newtext{to simply avoid possibly negative values of
  parameters generated for individual realizations. The mean values
  are taken again from~\cite{Sejnoha:MS:2008}.}

Geometrical model and boundary conditions used in numerical
calculations appear in Fig.~\ref{fig:EO1}(b). The finite element mesh
consisted of 662 triangular and 57 interface elements.  The calculated
curves entering the inverse analysis are plotted in
Fig.~\ref{fig:NC4}(a). The results representing the temperature in the
area of interface transition zone (ITZ) are displayed in
Fig.~\ref{fig:NC4}(b).  The numerically predicted temperature jumps
are presented in Fig.~\ref{fig:NC2}(a) assuming a single value of
parameter $\alpha_{\mathrm{int}}$ independent of both temperature and
flow direction.  The calculated and measured temperature profiles for
optimally fitted thermal data are compared in Fig.~\ref{fig:NC4}(b).

\begin{figure}[p]
\begin{center}
\begin{tabular}{cc}
\includegraphics*[width=65mm,keepaspectratio]{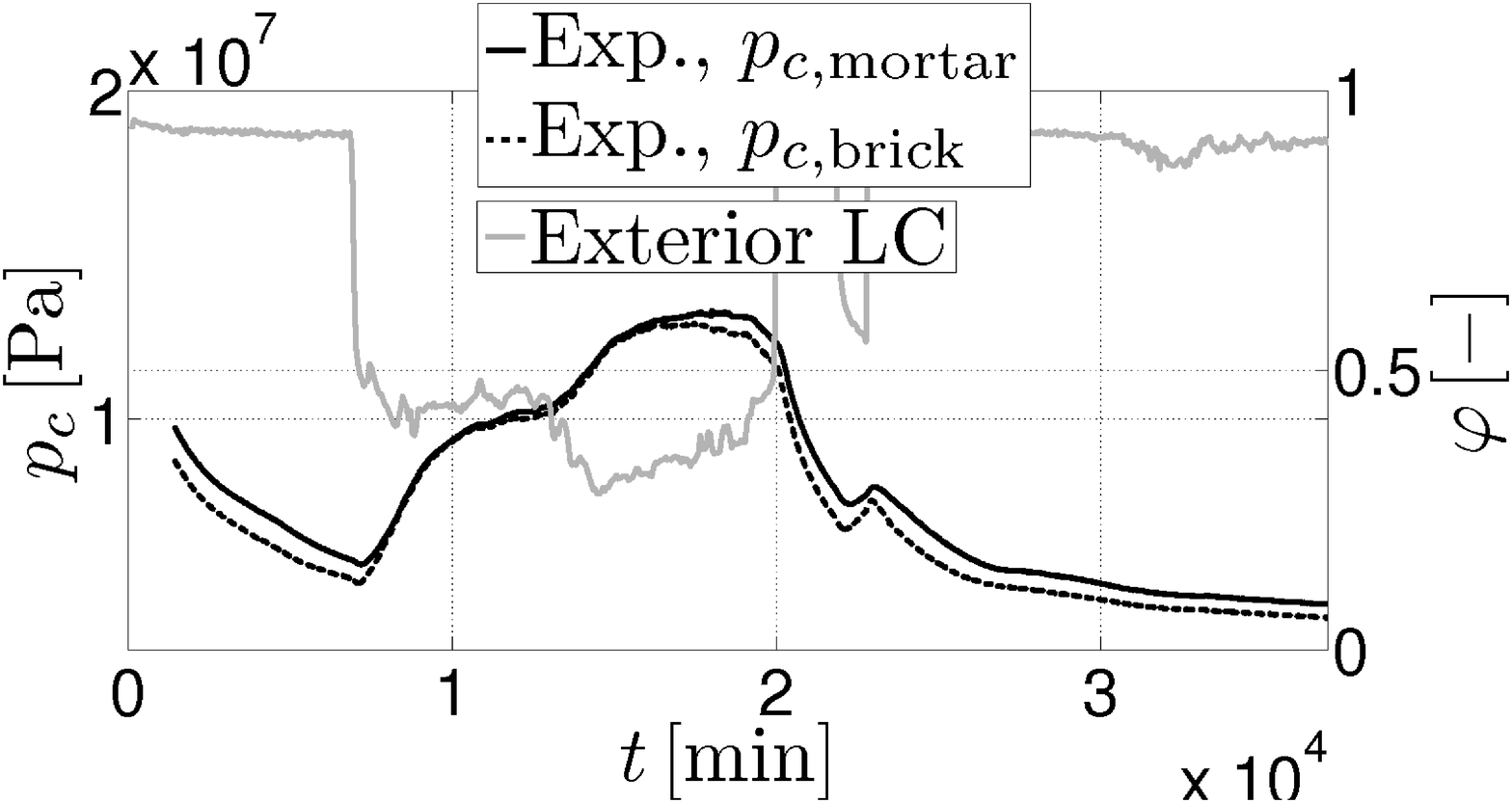}&
\includegraphics*[width=65mm,keepaspectratio]{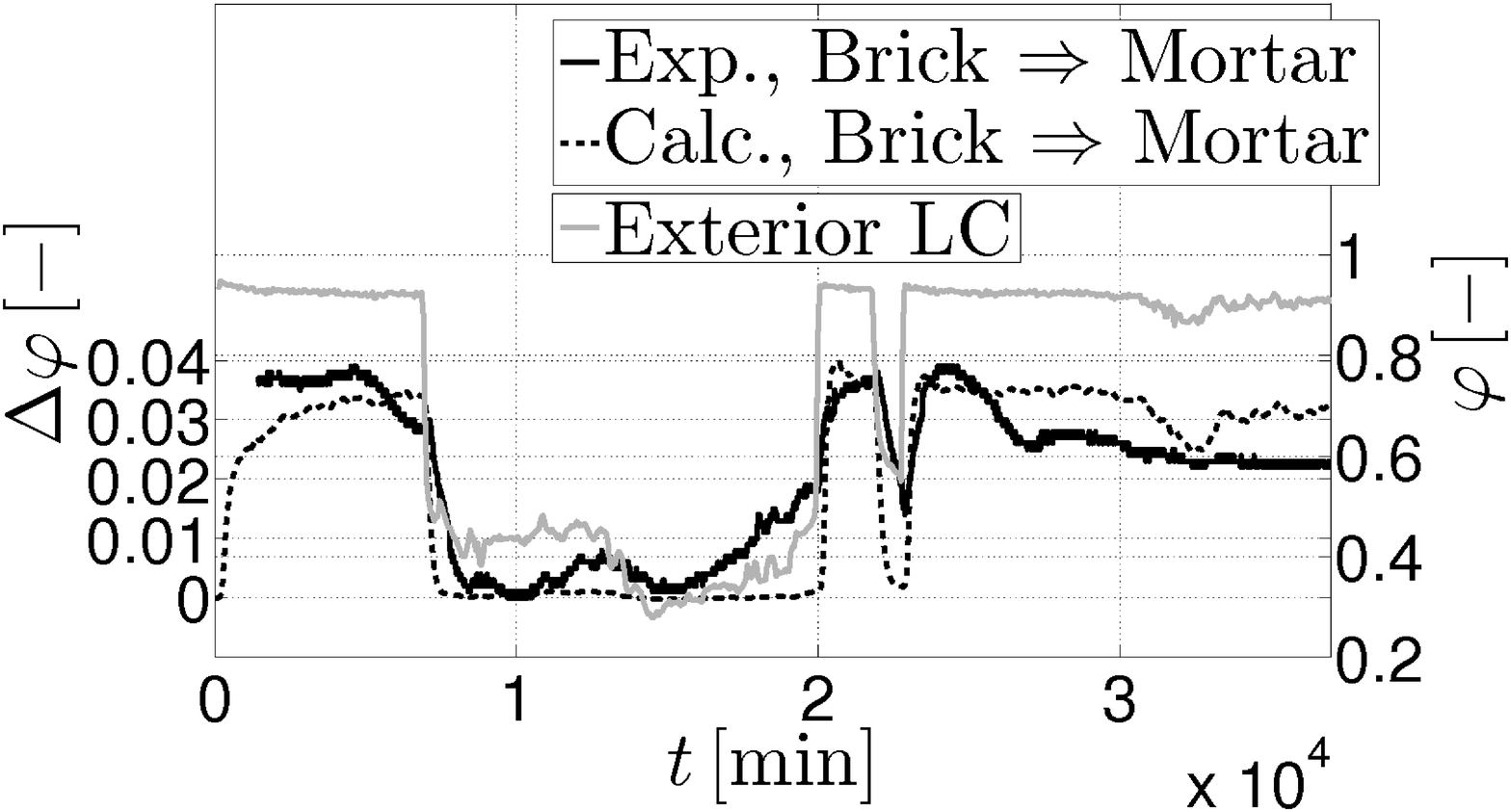}\\
(a) & (b)\\
\end{tabular}
\caption{Experiment No. 2: (a) distribution of back calculated
  capillary pressures along the interface, (b) comparison of
  experimentally and numerically obtained relative humidity jumps
  along the interface } \label{fig:NC9}
\end{center}
\begin{center}
\begin{tabular}{cc}
\includegraphics*[width=65mm,keepaspectratio]{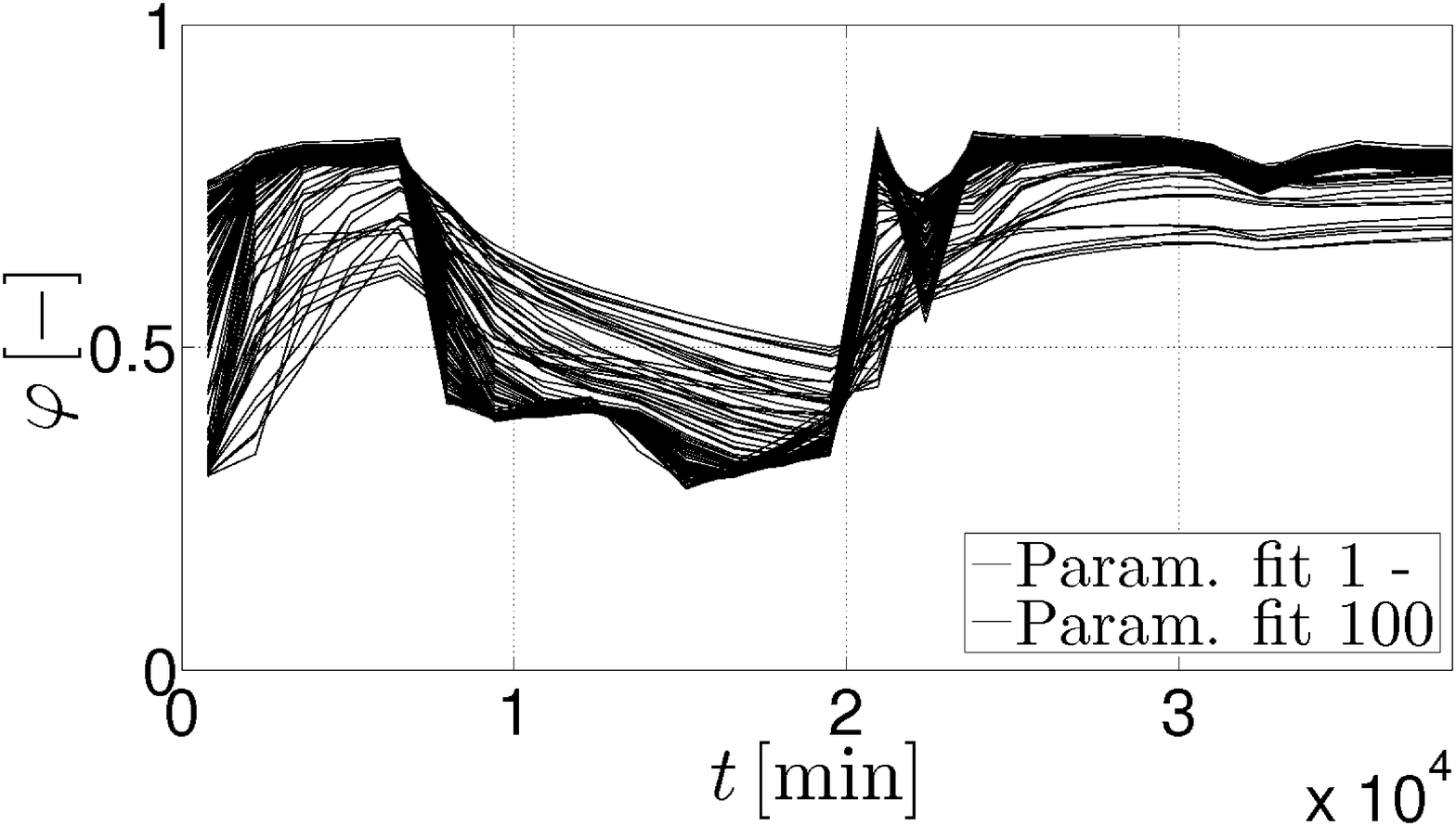}&
\includegraphics*[width=65mm,keepaspectratio]{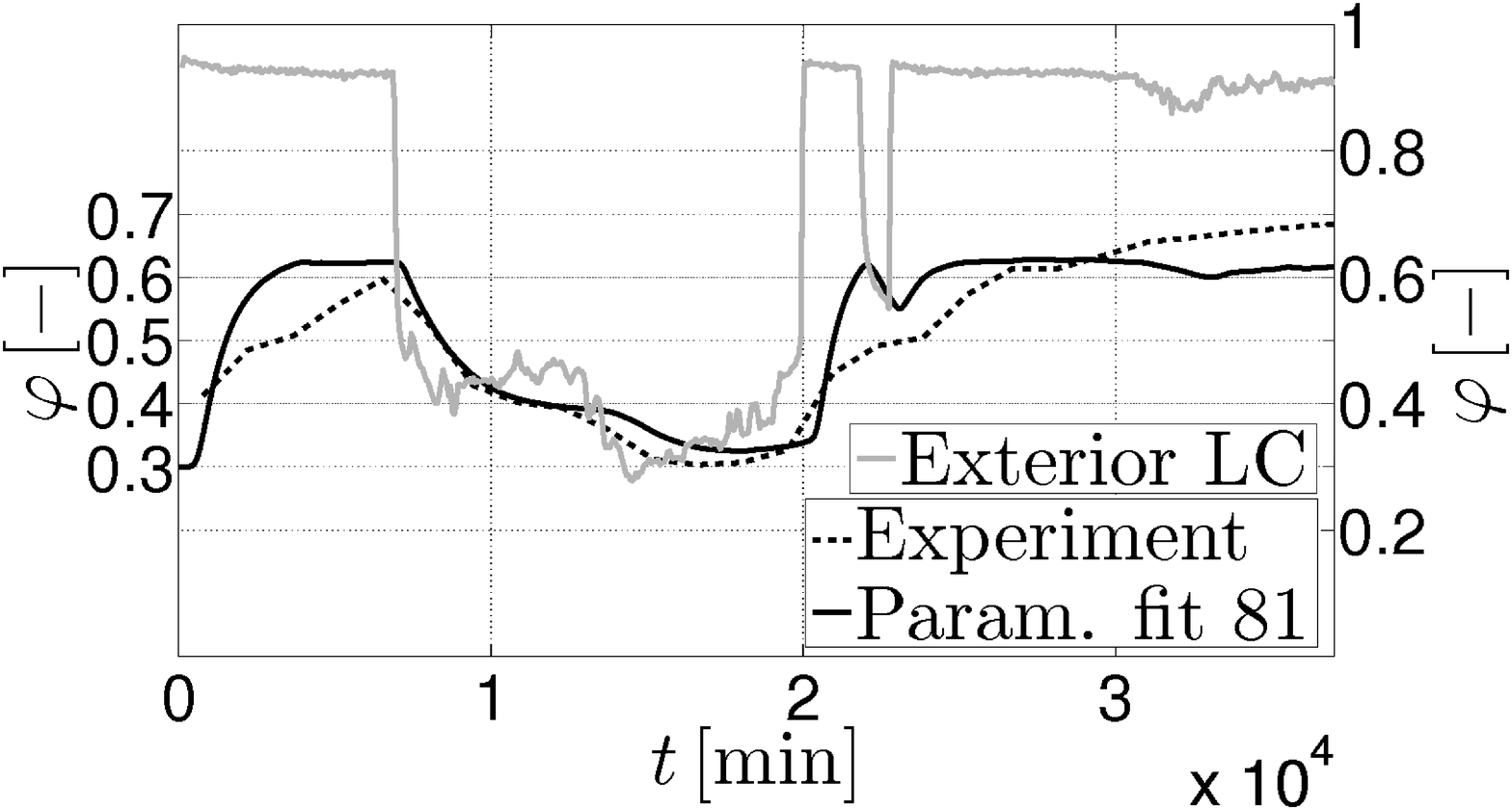}\\
(a) & (b)\\
\end{tabular}
\caption{Experiment No. 2: (a) evolution of relative humidity obtained
  from all numerical calculations in sensor No. $14$, (b) evolution of
  relative humidity obtained numerically for optimal data and from
  experiment in sensor No. $14$} \label{fig:NC8}
\end{center}
\begin{center}
\begin{tabular}{cc}
\includegraphics*[width=65mm,keepaspectratio]{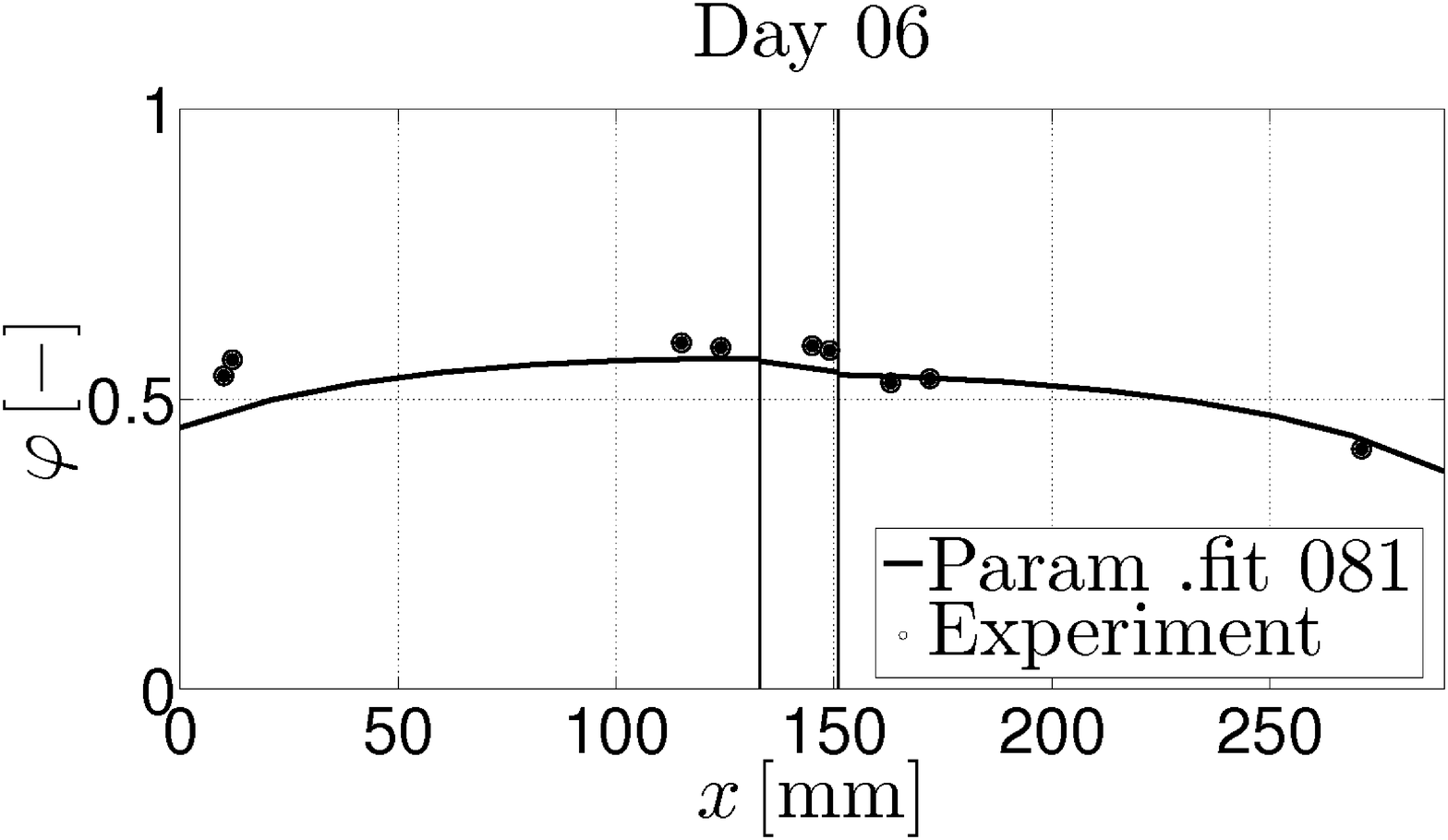}&
\includegraphics*[width=65mm,keepaspectratio]{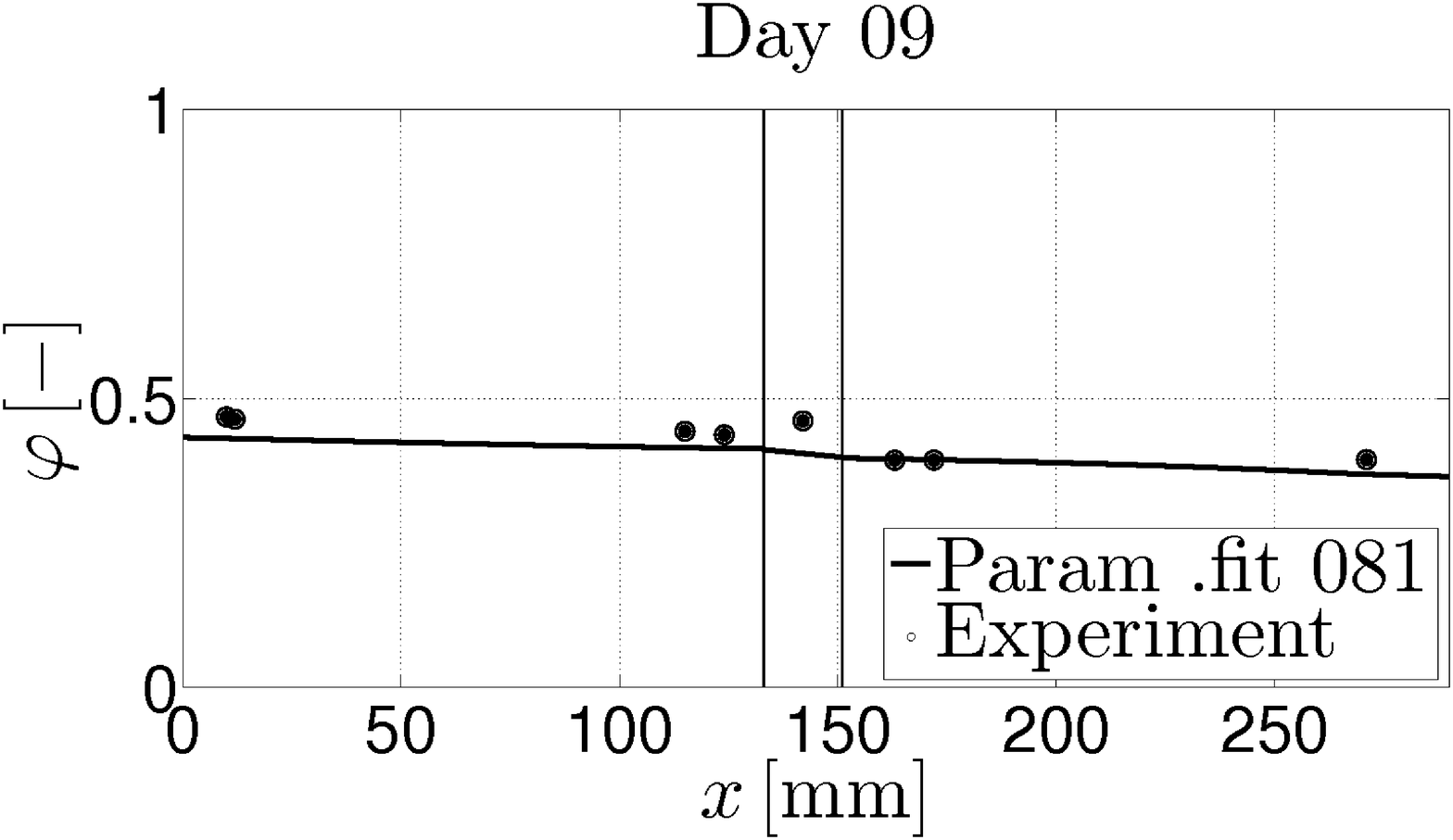}\\
(a) & (b)\\
\end{tabular}
\caption{Experiment No. 2: Calculated and measured relative humidity
  profile in chosen sensors corresponding to: (a) $6^\mathrm{th}$ day,
  (b) $9^\mathrm{th}$ day} \label{fig:NC7}
\end{center}
\end{figure}

The second experiment simulated a water vapor transport within a
steady state temperature profile. On the exterior side of the wall the
relative humidity of 0.3 [-] and on the interior side of 0.95 [-] were
prescribed, see Fig.~\ref{fig:NC5}. The temperature varied in the
range of 25 to 29 [$^{\circ}\mathrm{C}$]. This experiment lasted 27
days. \newtext{As before independence of jumps in relative humidity
  with respect to the flow direction was again observed.  Thus only
  jumps pertinent to brick-mortar transition are presented, see
  Fig.~\ref{fig:NC9}(b). Noticing that the actual values of these
  jumps are essentially comparable to the measuring precision may
  support the notion of a perfect hydraulic contact. In the next
  paragraphs we seek to confirm this via numerical simulations.}

As for numerical calculations, a similar generation of material
parameters as in the previous numerical example was carried out, see
Fig.~\ref{fig:NC8}, to capture the necessary relative humidity input
data for both material phases (free water saturation $w_f$
\newtext{[kgm$^{-3}$]}, water content $w_{80}$\newtext{[kgm$^{-3}$]}
at 0.8 [-] relative humidity, water vapor diffusion resistance factor
$\mu$ \newtext{[-]}, water absorption coefficient $A$
\newtext{[kgm$^{-2}$s$^{-0.5}$]} and interface permeability
$\beta_{\mathrm{int}}$ \newtext{[kgm$^{-2}$s$^{-1}$Pa$^{-1}$]}. Again,
the finite element mesh as seen in Fig.~\ref{fig:EO1}(b) was used with
the thermal material parameters of mortar and bricks
($\left\{\lambda_{0,\mathrm{m}},
\lambda_{0,\mathrm{b}}\right\}=\left\{0.45, 0.25\right\}$,
$\left\{b_{\mathrm{tcs,m}}, b_{\mathrm{tcs,b}}\right\}=\left\{9.0,
10.0\right\}$, $\alpha_{\mathrm{int}} = 100000$) found from the first
identification problem.

Fig.~\ref{fig:NC8}(a) represents a collection of all possible
variations of relative humidity, while Fig.~\ref{fig:NC8}(b) displays
the evolution of moisture for optimally fitted set of parameters
($\left\{w_{\mathrm{f,m}},
w_{\mathrm{f,b}}\right\}=\left\{160.0, 229.30\right\}$,
$\left\{w_{80,\mathrm{m}},
w_{80,\mathrm{b}}\right\}=\left\{22.72, 141.68\right\}$,
$\left\{\mu_{\mathrm{m}},
\mu_{\mathrm{b}}\right\}=\left\{9.63, 16.80\right\}$,
$\left\{A_{\mathrm{m}}, A_{\mathrm{b}}\right\}=\left\{0.82,
0.51\right\}$, $\beta_{\mathrm{int}} = 5.25\times{10}^{-9}$) derived from the "trial
and error" procedure.

The experimentally and numerically obtained results for the interface
transition zone are plotted in Fig.~\ref{fig:NC9} depicting
distributions of capillary pressures and jumps in relative humidity in
individual phases in the vicinity of the interface. Note that
capillary pressures are not directly measured but arise from back
calculation by introducing the corresponding values of experimentally
measured relative humidities in Kelvin's
equation~\eqref{jump_Kelvin}. Their discontinuous variation further
supports our assumption of an imperfect hydraulic contact. As seen
from both figures the experimental measurement was affected by the
fault of boundary condition on the exterior side of the masonry wall
due to electricity shutdown so that the identification of jumps in
this time period was rather complicated and therefore, as can be seen
from Fig.~\ref{fig:NC9}(b), the agreement between numerical and
experimental estimates is less accurate. It is also worth noting that
the jump magnitude in relative humidity is almost equal to the
accuracy of sensors. Therefore, incorporating the interface elements
into numerical calculations will probably have negligible influence on
the predicted results.

The calculated moisture profile and measured relative humidity for the
optimally fitted thermal and moisture material parameters are depicted
in Fig.~\ref{fig:NC7}.

\section{Homogenization on meso-scale}
\label{sec:HMS}
\begin{figure} [ht!]
\begin{center}
\begin{tabular}{c}
\includegraphics[width=50mm,keepaspectratio]{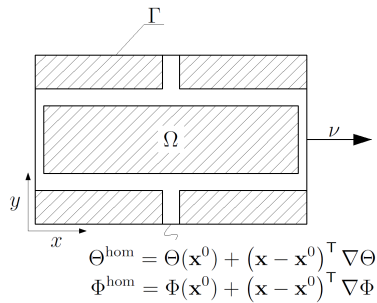}\\
\end{tabular}
\caption{Periodic unit cell of brick-mortar composite with assumed
  boundary conditions} \label{fig:RVR}
\end{center}
\end{figure}

Once having derived the material parameters of individual phases we
may now proceed to address the two principal objectives, recall the
introductory part, in the light of the first order homogenization
theory. To that end, we consider an RVE in terms of a periodic unit
cell (PUC) describing the geometrical and material details of the
meso-scale, see Fig.~\ref{fig:RVR}.

\subsection{Fundamentals of 1st order homogenization}
Since examining only the coupling effect and influence of interface
transition parameters on the homogenized properties, it is sufficient
to consider a steady state problem and perform a detailed parametric
study on meso-scale. Because only first order homogenization is
adopted, it is assumed that macroscopic temperature and relative
humidity vary only linearly over the PUC. This can be achieved by
loading the boundary of the PUC by the prescribed temperature
$\theta^{\mathrm{hom}}$ and relative humidity $\varphi^{\mathrm{hom}}$ derived
from uniform macroscopic temperature $\vek{\nabla}\Theta$ and relative
humidity $\vek{\nabla}\Phi$ gradients, respectively, as depicted in
Fig.~\ref{fig:RVR}.

In such a case, the local temperature and relative humidity
admit the following decomposition
\begin{eqnarray}
\theta(\vek{x}) & = & \Theta(\vek{x}^0) + \left(\vek{x} -
\vek{x}^{0}\right)^{\sf T}\vek{\nabla}\Theta+\theta^{*}(\vek{x})\,=\,\Theta^{\mathrm{hom}}+\theta^*, \nonumber\\
\varphi(\vek{x}) & = & \Phi(\vek{x}^0) + \left(\vek{x} -
\vek{x}^0\right)^{\sf T}\vek{\nabla}\Phi +\varphi^{*}(\vek{x})\,=\,\Phi^{\mathrm{hom}}+\varphi^*,
\label{eq:FH2}
\end{eqnarray}
where $\vek{\nabla}\Theta$ and $\vek{\nabla}\Phi$ are the
macroscopically uniform temperature and relative humidity gradients,
respectively. Fluctuations of local fields about the macroscopic ones
are denoted by $\theta^{*}(\vek{x})$ and
$\varphi^{*}(\vek{x})$. Finally, the temperature $\Theta(\vek{x}^0)$
and the relative humidity $\Phi(\vek{x}^0)$ at the reference point
$\vek{x}^0$ are introduced to uniquely define the distributions of the
corresponding local fields. The micro-temperature and micro-relative
humidity gradients
\begin{eqnarray}
\vek{\nabla}\theta(\vek{x}) & = &\vek{\nabla}\Theta+\vek{\nabla}\theta^{*}(\vek{x}),
\nonumber\\
\vek{\nabla}\varphi(\vek{x}) & = & \vek{\nabla}\Phi+\vek{\nabla}\varphi^{*}(\vek{x}), \label{eq:FH3}
\end{eqnarray}
averaged over the volume $|\Omega|$ of the PUC
\begin{eqnarray}
\langle \vek{\nabla}\theta(\vek{x}) \rangle & = &
\frac{1}{|\Omega|}\int_{\Omega} \vek{\nabla}\theta(\vek{x})
\mathrm{d}\Omega(\vek{x}) = \vek{\nabla}\Theta +
\frac{1}{|\Omega|}\int_{\Omega} \vek{\nabla}\theta^{*}(\vek{x})
\mathrm{d}\Omega(\vek{x}),\nonumber \\
\langle \vek{\nabla}\varphi(\vek{x}) \rangle & = &
\frac{1}{|\Omega|}\int_{\Omega} \vek{\nabla}\varphi(\vek{x})
\mathrm{d}\Omega(\vek{x}) = \vek{\nabla} \Phi +
\frac{1}{|\Omega|}\int_{\Omega} \vek{\nabla}\varphi^{*}(\vek{x})
\mathrm{d}\Omega(\vek{x}),
\end{eqnarray}
yield the scale transition relation, see e.g.~\cite{Ozdemir:IJNME:2008},
\begin{eqnarray}
\langle \vek{\nabla}\theta^{*} \rangle &=&
\frac{1}{|\Omega|}\int_{\Omega} \vek{\nabla}\theta^{*}(\vek{x})
\mathrm{d}\Omega(\vek{x}) = \frac{1}{|\Omega|}\int_{\Gamma}
\theta^{*}(\vek{x})\vek{\nu} \mathrm{d}\Gamma(\vek{x}) = 0,
\nonumber \\
\langle \vek{\nabla}\varphi^{*} \rangle &=&
\frac{1}{|\Omega|}\int_{\Omega} \vek{\nabla}\varphi^{*}(\vek{x})
\mathrm{d}\Omega(\vek{x}) = \frac{1}{|\Omega|}\int_{\Gamma}
\varphi^{*}(\vek{x})\vek{\nu} \mathrm{d}\Gamma(\vek{x}) = 0.
\end{eqnarray}
The boundary integral disappears providing either the fluctuation
parts of the local fields equal zero (Dirichlet boundary conditions)
or the periodic boundary conditions, i.e. the same values of
$\theta^*$ and $\varphi^{*}$ on opposite sides of the rectangular PUC,
are enforced on $\Gamma$.

\subsection{Discretized form of energy balance equation}
Using Eqs.~\eqref{TFlux-1} and~\eqref{TFlux-3} the heat flux density
is given by
\begin{eqnarray}
\vek{q} + h_{v}\vek{g}_{v} & = & -\lambda\vek{\nabla}\theta
- h_{v}\delta_{p}\vek{\nabla}(\varphi p_{\mathrm{sat}})\nonumber \\[0.3cm]
& = & -\lambda\vek{\nabla} \theta - h_{v}\delta_{p}
p_{\mathrm{sat}} \vek{\nabla}\varphi - h_{v}\delta_{p}
\varphi \vek{\nabla}p_{\mathrm{sat}}. \label{eq:AHF1}
\end{eqnarray}
Next, substituting for local fields $\theta$ and $\varphi$ from
Eq.~\eqref{FH3} into \eqref{AHF1} and after some manipulations we get
\begin{equation}
\vek{q} + h_{v}\vek{g}_{v}
 =  - [\lambda + h_{v}\delta_{p} \frac{\mathrm{d}
p_{\mathrm{sat}}}{\mathrm{d}\theta}\varphi](\vek{\nabla}\Theta +
\vek{\nabla}\theta^{*})-h_{v}\delta_{p}
p_{\mathrm{sat}}(\vek{\nabla}\Phi + \vek{\nabla}\varphi^{*}).\label{eq:AHF2}
\end{equation}

The stepping stone in the estimation of macroscopic response is the
Hill-Mandel lemma suggesting equality of the work of local fields
averaged over the solution domain and the work of their macroscopic
counterparts. Owing to the nonlinear nature of underlying material
models, the formulation must be written in an incremental form. To
that end, assume an equilibrium state at the end of the $i$-th time
step and consider a small increment of the macroscopic temperature
gradient $\mathrm{d}(\vek{\nabla}\Theta)$ resulting in an incremental
change of local and macroscopic fluxes so that
\begin{equation}
\avgs{-\delta(\vek{\nabla}\theta)\trn\mathrm{d}(\vek{q} +h_{v}\vek{g}_{v})}=
-\delta(\vek{\nabla}\Theta)\trn\mathrm{d}(\vek{q}^{\mathrm{M}}
+h_{v}^{\mathrm{M}}\vek{g}_{v}^{\mathrm{M}}),\label{eq:AHF3}
\end{equation}
where the symbol
$\langle\cdot\rangle=\frac{1}{|\Omega|}\int_{\Omega}\cdot\mathrm{d}\Omega$
represents the volume averaging. Providing we admit only the
Dirichlet boundary conditions in the form of the prescribed
gradient of macroscopic temperature the term on the right hand
side would disappear ($\delta(\vek{\nabla}\Theta)=\tenss{0}$). For
the calculation purposes this condition is considered to determine
the unknown fluctuation fields. Nevertheless, for the derivation
of instantaneous macroscopic conductivity matrix it will prove
useful to keep the full generality of the mathematical
formulation.

To proceed, consider the numerical solution in the framework of the
finite element method and introduce the standard geometric matrix
$\tensf{B}$ storing the spatial derivatives of the finite element
shape functions to get the discretized form of the local fields as
\begin{equation}
\vek{\nabla}\theta=\vek{\nabla}\Theta+\tensf{B}\vek{r}_{\theta}^{*},\qquad
\vek{\nabla}\varphi=\vek{\nabla}\Phi+\tensf{B}\vek{r}_{\varphi}^{*},\label{eq:AHF4}
\end{equation}
and
\begin{equation}
\delta(\vek{\nabla}\theta)=\delta(\vek{\nabla}\Theta)+\tensf{B}\delta\vek{r}_{\theta}^{*},\qquad
\delta(\vek{\nabla}\varphi)=\delta(\vek{\nabla}\Phi)+\tensf{B}\delta\vek{r}_{\varphi}^{*}.\label{eq:AHF5}
\end{equation}
Substituting Eqs.~\eqref{AHF2},~\eqref{AHF4} and \eqref{AHF5} into
Eq.~\eqref{AHF3} provides
\begin{eqnarray}
&&\delta(\vek{\nabla}\Theta)\trn\{
\mathrm{d}(\vek{q}^{\mathrm{M}}+h_{v}^{\mathrm{M}}\vek{g}_{v}^{\mathrm{M}}) + \nonumber \\[0.3cm]
&&\qquad +
\underbrace{\avgs{\lambda + h_{v}\delta_{p}\frac{\mathrm{d} p_{\mathrm{sat}}}{\mathrm{d}
\theta}\hat{\Phi}^{\mathrm{hom}}}}_{\tensf{K}_{\theta\theta}^{\mathrm{m}}}\mathrm{d}(\vek{\nabla}\Theta) +
\underbrace{\avgs{h_{v}\delta_{p} p_{\mathrm{sat}}}}_{\tensf{K}_{\theta\varphi}^{\mathrm{m}}}\mathrm{d}(\vek{\nabla}\Phi) +
\nonumber \\[0.3cm]
&&\qquad +
\underbrace{\avgs{[\lambda +
h_{v}\delta_{p}\frac{\mathrm{d} p_{\mathrm{sat}}}{\mathrm{d}
\theta}\hat{\Phi}^{\mathrm{hom}}]\tensf{B}}}_{\tensf{L}_{\theta\theta}}\mathrm{d}\vek{r}_{\theta}^{*}
+
\underbrace{\avgs{h_{v}\delta_{p}p_{\mathrm{sat}}\tensf{B}}}_{\tensf{L}_{\theta\varphi}} \mathrm{d}\vek{r}_{\varphi}^{*}\}
+ \nonumber \\[0.3cm]
&&+\delta(\vek{r}_{\theta}^{*})\trn\{
\underbrace{\avgs{\tensf{B}^{\sf T} [\lambda + h_{v}\delta_{p}
\frac{\mathrm{d} p_{\mathrm{sat}}}{\mathrm{d}\theta}\hat{\Phi}^{\mathrm{hom}}]}}_{\tensf{L}_{\theta\theta}^{\sf T}}
\mathrm{d}(\vek{\nabla}\Theta) +
\underbrace{\avgs{\tensf{B}^{\sf T}h_{v}\delta_{p} p_{\mathrm{sat}}}}_{\tensf{L}_{\theta\varphi}^{\sf T}}\mathrm{d}(\vek{\nabla}\Phi) +
\nonumber \\[0.3cm]
&&\qquad +
\underbrace{\avgs{\tensf{B}^{\sf T} [\lambda +
h_{v}\delta_{p}\frac{\mathrm{d} p_{\mathrm{sat}}}{\mathrm{d}
\theta}\hat{\Phi}^{\mathrm{hom}}]\tensf{B}}}_{\tensf{K}_{\theta\theta}}
\mathrm{d}\vek{r}_{\theta}^{*} +
\underbrace{\avgs{\tensf{B}\trn
h_{v}\delta_{p} p_{\mathrm{sat}}
\tensf{B}}}_{\tensf{K}_{\theta\varphi}}
\mathrm{d}\vek{r}_{\varphi}^{*}\}
= 0, \label{eq:AHF6}
\end{eqnarray}
where the notation
$\hat{\Phi}^{\mathrm{hom}}={\Phi}^{\mathrm{hom}}+\hat{\varphi}^*$ is
introduced to identify $\hat{\varphi}^*$ with the fluctuation of
relative humidity at the end of the previous iteration step.
Because of the independence of variations $\delta(\vek{\nabla}\Theta)$ and
$\delta\vek{r}_{\theta}^{*}$, Eq.~\eqref{AHF6} splits into two
equalities. Noting that the change of the macroscopic flux can be expressed in
the form
\begin{equation}
\mathrm{d}(\vek{q}^{\mathrm{M}}
+h_{v}^{\mathrm{M}}\mathrm{d}\vek{g}_{v}^{\mathrm{M}})=-\tensf{K}_{\theta\theta}^{\mathrm{M}}\mathrm{d}(\vek{\nabla}\Theta)-
\tensf{K}_{\theta\varphi}^{\mathrm{M}}\mathrm{d}(\vek{\nabla}\Phi),\label{eq:AHF8}
\end{equation}
we write the first equality as
\begin{equation}
(-\tensf{K}_{\theta\theta}^{\mathrm{M}}+\tensf{K}_{\theta\theta}^{\mathrm{m}})\mathrm{d}(\vek{\nabla}\Theta)
+(-\tensf{K}_{\theta\varphi}^{\mathrm{M}}+\tensf{K}_{\theta\varphi}^{\mathrm{m}})\mathrm{d}(\vek{\nabla}\Phi)
+\tensf{L}_{\theta\theta}\mathrm{d}\vek{r}_{\theta}^{*} +
\tensf{L}_{\theta\varphi}\mathrm{d}\vek{r}_{\varphi}^{*}=0.\label{eq:AHF9}
\end{equation}
The second equality is then provided by
\begin{equation}
\tensf{L}_{\theta\theta}^{\sf T}\mathrm{d}(\vek{\nabla}\Theta)+\tensf{L}_{\theta\varphi}^{\sf T}\mathrm{d}(\vek{\nabla}\Phi)
+\tensf{K}_{\theta\theta}\mathrm{d}\vek{r}_{\theta}^{*}+
\tensf{K}_{\theta\varphi}\mathrm{d}\vek{r}_{\Phi}^{*}=0.\label{eq:AHF10}
\end{equation}

\subsection{Discretized form of mass balance equation}
Similar to Eq.~\eqref{AHF3}) the variational form of the Hill-Mandel
lemma for moisture transport reads
\begin{equation}
\avgs{-\delta(\vek{\nabla}\varphi)\trn[\mathrm{d}(\vek{g}_{v} +\vek{g}_{w})]} =
-\delta(\vek{\nabla}\Phi)^{\sf T}[\vek{g}_{v}^{\mathrm{M}}\mathrm{d}(\vek{g}_{v}^{\mathrm{M}}+\vek{g}_{w}^{\mathrm{M}})],\label{eq:AMF1}
\end{equation}
where the moisture flux density is given by
\begin{equation}
\vek{g}_{v}+\vek{g}_{w} =  -(D_{\varphi}+
\delta_{p}p_{\mathrm{sat}})(\vek{\nabla}\Phi +
\vek{\nabla}\varphi^{*}) - \delta_{p} \frac{\mathrm{d}
p_{\mathrm{sat}}}{\mathrm{d} \theta}\varphi(\vek{\nabla}\Theta +
\vek{\nabla}\theta^{*}).\label{eq:AMF2}
\end{equation}
Introducing Eq.~\eqref{AMF2} together with Eqs.~\eqref{AHF4} and
\eqref{AHF5} into Eq.~\eqref{AMF1} yields
\begin{eqnarray}
&&\delta(\vek{\nabla}\Phi)\trn\{\mathrm{d}(\vek{g}_{v}^{\mathrm{M}}
+\vek{g}_{w}^{\mathrm{M}}) + \nonumber \\[0.3cm]
&&\qquad +
\underbrace{\avgs{\delta_{p} \frac{\mathrm{d}p_{\mathrm{sat}}}{\mathrm{d}\theta}\hat{\Phi}^{\mathrm{hom}}}}_{\tensf{K}_{\varphi\theta}^{\mathrm{m}}}
\mathrm{d}(\vek{\nabla}\Theta) +
\underbrace{\avgs{D_{\varphi}+\delta_{p}p_{\mathrm{sat}}}}_{\tensf{K}_{\varphi\varphi}^{\mathrm{m}}}
\mathrm{d}(\nabla\Phi) + \nonumber \\[0.3cm]
&&\qquad +
\underbrace{\avgs{[\delta_{p} \frac{\mathrm{d}p_{\mathrm{sat}}}{\mathrm{d}\theta}\hat{\Phi}^{\mathrm{hom}}]\tensf{B}}}_{\tensf{L}_{\varphi\theta}}\mathrm{d}\vek{r}_{\theta}^{*}
+ \underbrace{\avgs{[D_{\varphi}+\delta_{p}p_{\mathrm{sat}}]\tensf{B}}}_{\tensf{L}_{\varphi\varphi}}
\mathrm{d}\vek{r}_{\varphi}^{*}\}
+ \nonumber \\[0.3cm]
&&+\delta(\vek{r}_{\varphi}^{*})\trn
\{\underbrace{\avgs{\tensf{B}^{\sf T} [\delta_{p} \frac{\mathrm{d}p_{\mathrm{sat}}}{\mathrm{d}\theta}\hat{\Phi}^{\mathrm{hom}}]}}_{\tensf{L}_{\varphi\theta}^{\sf T}}
\mathrm{d}(\vek{\nabla}\Theta) +
\underbrace{\avgs{\tensf{B}^{\sf T}[D_{\varphi}+\delta_{p}p_{\mathrm{sat}}]}}_{\tensf{L}_{\varphi\varphi}^{\sf T}}\mathrm{d}(\vek{\nabla}\Phi)
+ \nonumber \\[0.3cm]
&&\qquad +
\underbrace{\avgs{\tensf{B}^{\sf T} [\delta_{p}\frac{\mathrm{d} p_{\mathrm{sat}}}{\mathrm{d}\theta}\hat{\Phi}^{\mathrm{hom}}]\tensf{B}}}_{\tensf{K}_{\varphi\theta}}
\mathrm{d}\vek{r}_{\theta}^{*} +
\underbrace{\avgs{\tensf{B}^{\sf T}[D_{\varphi}+\tensf{\delta}_{p}p_{\mathrm{sat}}]\tensf{B}}}_{\tensf{K}_{\varphi\varphi}}
\mathrm{d}\vek{r}_{\varphi}^{*}\}=0.
\end{eqnarray}
Since (compare with~\eqref{AHF8})
\begin{equation}
\mathrm{d}(\vek{g}_{v}^{\mathrm{M}}
+\vek{g}_{w}^{\mathrm{M}})=-\tensf{K}_{\varphi\theta}^{\mathrm{M}}\mathrm{d}(\vek{\nabla}\Theta)-
\tensf{K}_{\varphi\varphi}^{\mathrm{M}}\mathrm{d}(\vek{\nabla}\Phi),
\end{equation}
the Hill-Mandel lemma for moisture transport (compare with
Eq.~\eqref{AHF9}) becomes
\begin{equation}
(-\tensf{K}_{\varphi\theta}^{\mathrm{M}}+\tensf{K}_{\varphi\theta}^{\mathrm{m}})\mathrm{d}(\vek{\nabla}\Theta)
+(-\tensf{K}_{\varphi\varphi}^{\mathrm{M}}+\tensf{K}_{\varphi\varphi}^{\mathrm{m}})\mathrm{d}(\vek{\nabla}\Phi)
+\tensf{L}_{\varphi\theta}\mathrm{d}\vek{r}_{\theta}^{*} +
\tensf{L}_{\varphi\varphi}\mathrm{d}\vek{r}_{\varphi}^{*} =
0,\label{eq:AMF3}
\end{equation}
together with (recall Eg.~\eqref{AHF10})
\begin{equation}
\tensf{L}_{\varphi\theta}^{\sf T}\mathrm{d}(\vek{\nabla}\Theta)+\tensf{L}_{\varphi\varphi}^{\sf T}\mathrm{d}(\vek{\nabla}\Phi)
+\tensf{K}_{\varphi\theta}\mathrm{d}\vek{r}_{\theta}^{*}+
\tensf{K}_{\varphi\varphi}\mathrm{d}\vek{r}_{\varphi}^{*}=0.\label{eq:AMF4}
\end{equation}

\subsection{Macroscopic conductivity matrices}
Assuming that the heat and moisture balance is reached at the end
of the $i$-th step at the mesostructural level and the nodal
unbalanced forces are equal to zero, we obtain from the system of
equations~\eqref{AHF10} and \eqref{AMF4}
\begin{equation}
\left\{\begin{array}{c} \mathrm{d}\vek{r}_{\theta}^{*} \\
\mathrm{d}\vek{r}_{\varphi}^{*}\end{array}\right\} = -
\left[\begin{array}{cc} \tensf{K}_{\theta\theta} &
\tensf{K}_{\theta\varphi} \\ \tensf{K}_{\varphi\theta} &
\tensf{K}_{\varphi\varphi}\end{array}\right]^{-1}
\left[\begin{array}{cc} \tensf{L}_{\theta\theta}^{\sf T} &
\tensf{L}_{\theta\varphi}^{\sf T} \\ \tensf{L}_{\varphi\theta}^{\sf T} &
\tensf{L}_{\varphi\varphi}^{\sf T}\end{array}\right]
\left\{\begin{array}{c} \mathrm{d}(\vek{\nabla}\Theta) \\
\mathrm{d}(\vek{\nabla}\Phi)\end{array}\right\}. \label{eq:SSE1}
\end{equation}
The system of equations~\eqref{AHF9} and~\eqref{AMF1} can be rewritten
with help of Eq.~\eqref{SSE1} as
\begin{eqnarray}
&&\left[\begin{array}{cc}
\tensf{K}_{\theta\theta}^{\mathrm{M}} &
\tensf{K}_{\theta\varphi}^{\mathrm{M}} \\
\tensf{K}_{\varphi\theta}^{\mathrm{M}} &
\tensf{K}_{\varphi\varphi}^{\mathrm{M}}\end{array}\right]\left\{\begin{array}{c} \mathrm{d}(\vek{\nabla}\Theta) \\
\mathrm{d}(\vek{\nabla}\Phi)\end{array}\right\} =\nonumber \\
&&=\left(\left[\begin{array}{cc}
\tensf{K}_{\theta\theta}^{\mathrm{m}}
&\tensf{K}_{\theta\varphi}^{\mathrm{m}} \\
\tensf{K}_{\varphi\theta}^{\mathrm{m}} &
\tensf{K}_{\varphi\varphi}^{\mathrm{m}}\end{array}\right]-
\left[\begin{array}{cc}
\tensf{L}_{\theta\theta} & \tensf{L}_{\theta\varphi} \\
\tensf{L}_{\varphi\theta} &
\tensf{L}_{\varphi\varphi}\end{array}\right] \left[\begin{array}{cc}
\tensf{K}_{\theta\theta} & \tensf{K}_{\theta\varphi} \\
\tensf{K}_{\varphi\theta} &
\tensf{K}_{\varphi\varphi}\end{array}\right]^{-1}
\left[\begin{array}{cc} \tensf{L}_{\theta\theta}^{\sf T} &
\tensf{L}_{\theta\varphi}^{\sf T} \\ \tensf{L}_{\varphi\theta}^{\sf T} &
\tensf{L}_{\varphi\varphi}^{\sf T}\end{array}\right]\right)
\left\{\begin{array}{c} \mathrm{d}(\vek{\nabla}\Theta) \\
\mathrm{d}(\vek{\nabla}\Phi)\end{array}\right\},\nonumber
\end{eqnarray}
to finally arrive at the macroscopic conductivity matrix in the form
\begin{equation}
\left[\begin{array}{cc}
\tensf{K}_{\theta\theta}^{\mathrm{M}} &
\tensf{K}_{\theta\varphi}^{\mathrm{M}} \\
\tensf{K}_{\varphi\theta}^{\mathrm{M}} &
\tensf{K}_{\varphi\varphi}^{\mathrm{M}}\end{array}\right]
=\left[\begin{array}{cc}
\tensf{K}_{\theta\theta}^{\mathrm{m}}
&\tensf{K}_{\theta\varphi}^{\mathrm{m}} \\
\tensf{K}_{\varphi\theta}^{\mathrm{m}} &
\tensf{K}_{\varphi\varphi}^{\mathrm{m}}\end{array}\right]-
\left[\begin{array}{cc}
\tensf{L}_{\theta\theta} & \tensf{L}_{\theta\varphi} \\
\tensf{L}_{\varphi\theta} &
\tensf{L}_{\varphi\varphi}\end{array}\right] \left[\begin{array}{cc}
\tensf{K}_{\theta\theta} & \tensf{K}_{\theta\varphi} \\
\tensf{K}_{\varphi\theta} &
\tensf{K}_{\varphi\varphi}\end{array}\right]^{-1}
\left[\begin{array}{cc} \tensf{L}_{\theta\theta}^{\sf T} &
\tensf{L}_{\theta\varphi}^{\sf T} \\ \tensf{L}_{\varphi\theta}^{\sf T} &
\tensf{L}_{\varphi\varphi}^{\sf T}\end{array}\right].\label{eq:HOM}
\end{equation}

\subsection{Numerical results}
Recall the introductory part posting two particular issues to be
investigated. First, we were interested in assessing the influence of
interface transition parameters on the homogenized response.  For this
purpose, the PUC displayed in Fig.~\ref{fig:RVR} was loaded by a set
of prescribed macroscopic temperature $\vek{\nabla}\Theta$ and
relative humidity $\vek{\nabla}\Phi$ gradients. The initial values of
temperature and relative humidity were set equal to
$\Theta(\vek{x}^0)=20$ [$^{\circ}\mathrm{C}$] and
$\Phi(\vek{x}^0)=0.5$ [-], respectively. The model parameters were
found using equations listed in~\cite{Kunzel:IJHMT:1997} together with
phase material data derived in Section~\ref{sec:IP}. To see the
sensitivity of the homogenized properties on interface parameters, we
assumed a set of constant values of parameters $\alpha_{\mathrm{int}}$
and $\beta_{\mathrm{int}}$ slightly varying about the optimal ones
$\alpha_{\mathrm{int}}=10^5$ [Wm$^{-2}$K$^{-1}$] and
$\beta_{\mathrm{int}}=5.25\times{10}^{-9}$
[kgm$^{-2}$s$^{-1}$Pa$^{-1}$].

\begin{figure}[ht]
\begin{center}
\begin{tabular}{cc}
\includegraphics[width=57mm,keepaspectratio]{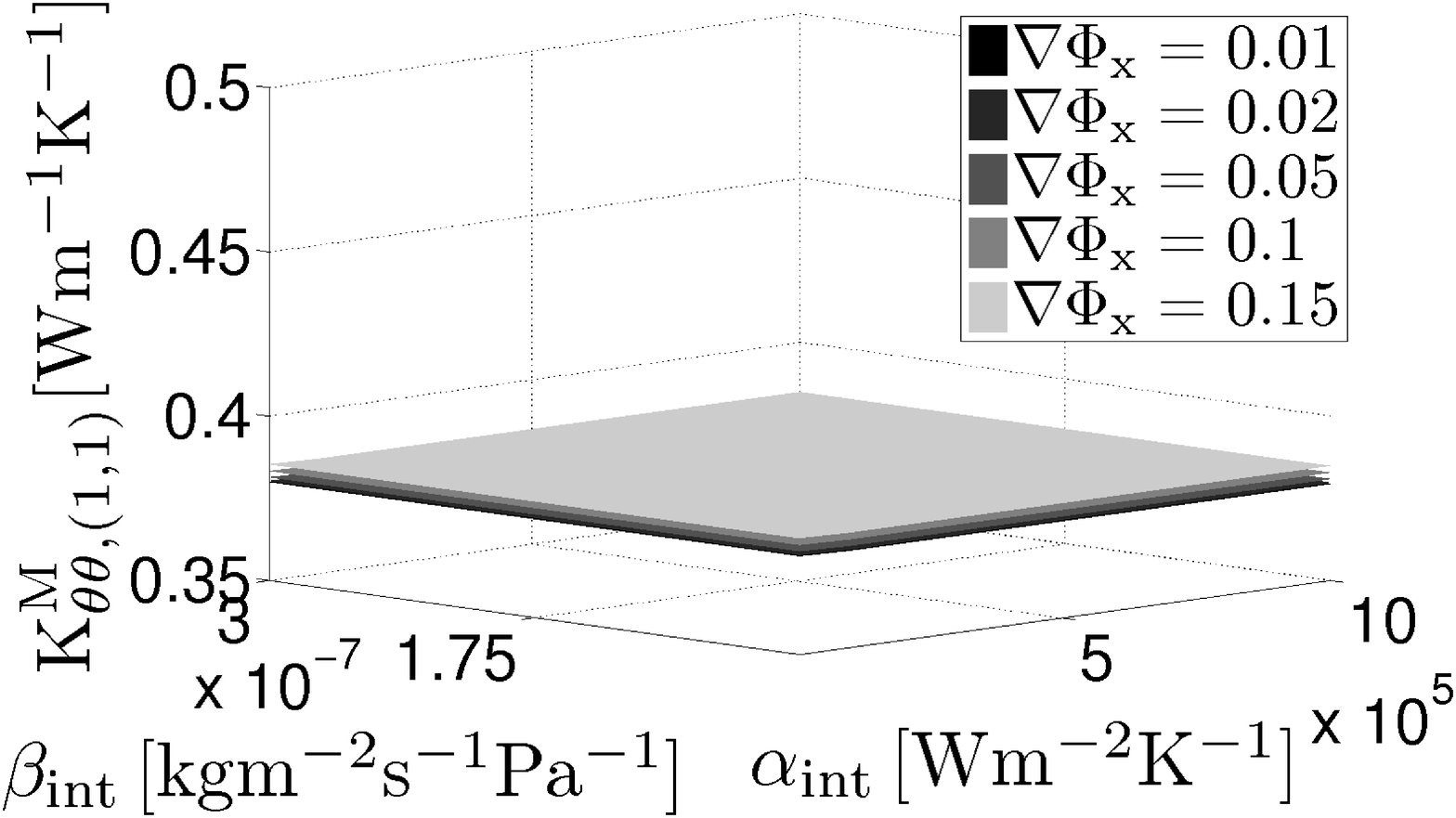}
& \includegraphics[width=57mm,keepaspectratio]{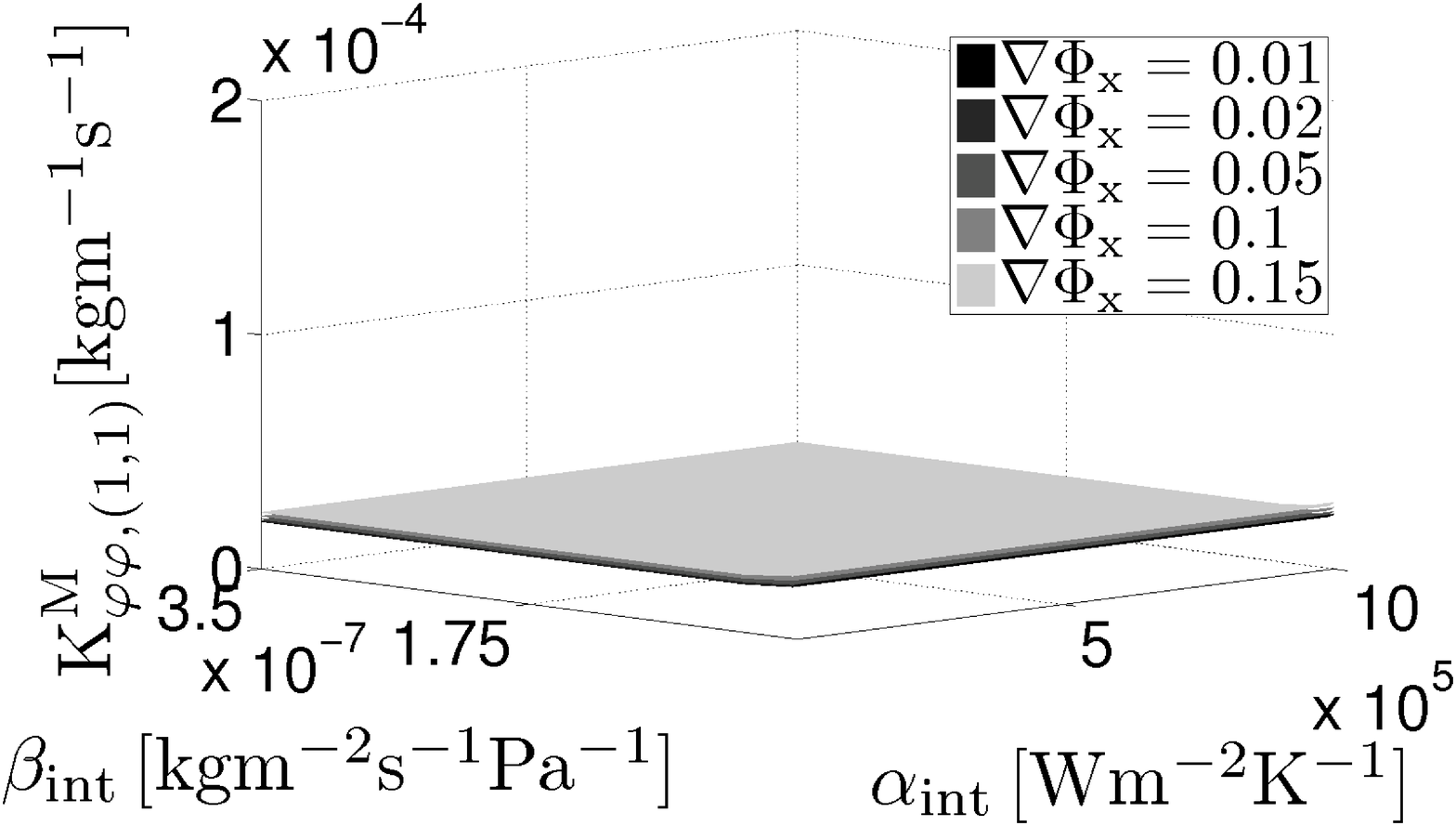}\\
(a) & (b)\\
\end{tabular}
\caption{Variation of the effective parameters
  ($\mathrm{K}_{\theta\theta,(1,1)}^{\mathrm{M}}$ and
  $\mathrm{K}_{\varphi\varphi,(1,1)}^{\mathrm{M}}$) as a function of
  the interface transfer coefficients for various values of relative
  humidity gradients} \label{fig:Rint}
\end{center}
\end{figure}

The results are plotted in Fig.~\ref{fig:Rint} showing variation of
the selected diagonal terms of the homogenized conductivity matrix,
recall Eq.~\eqref{HOM}. Clearly, while we may notice dependence of
homogenized terms on macroscopic gradients, the impact of variation of
$\alpha_{\mathrm{int}}$ and $\beta_{\mathrm{int}}$ is almost
imperceptible. These results together with experimental observations
presented in Section~\ref{sec:IP} may suggest that the introduction of
interface transition zone at the mesostructural level is essentially
negligible for the prediction of effective properties, at least for
the present material system and applied range of temperatures and
relative humidities.

Second, to capture the influence of macroscopic loading conditions on
effective parameters we assumed, in view of the previous results, a
perfect hydraulic contact and loaded the PUC again by several
different macroscopic temperature and relative humidity gradients. In
addition, the macroscopic/initial temperature and macroscopic/initial
relative humidity also varied.

\begin{figure}[ht]
\begin{center}
\begin{tabular}{cc}
\includegraphics[width=57mm,keepaspectratio]{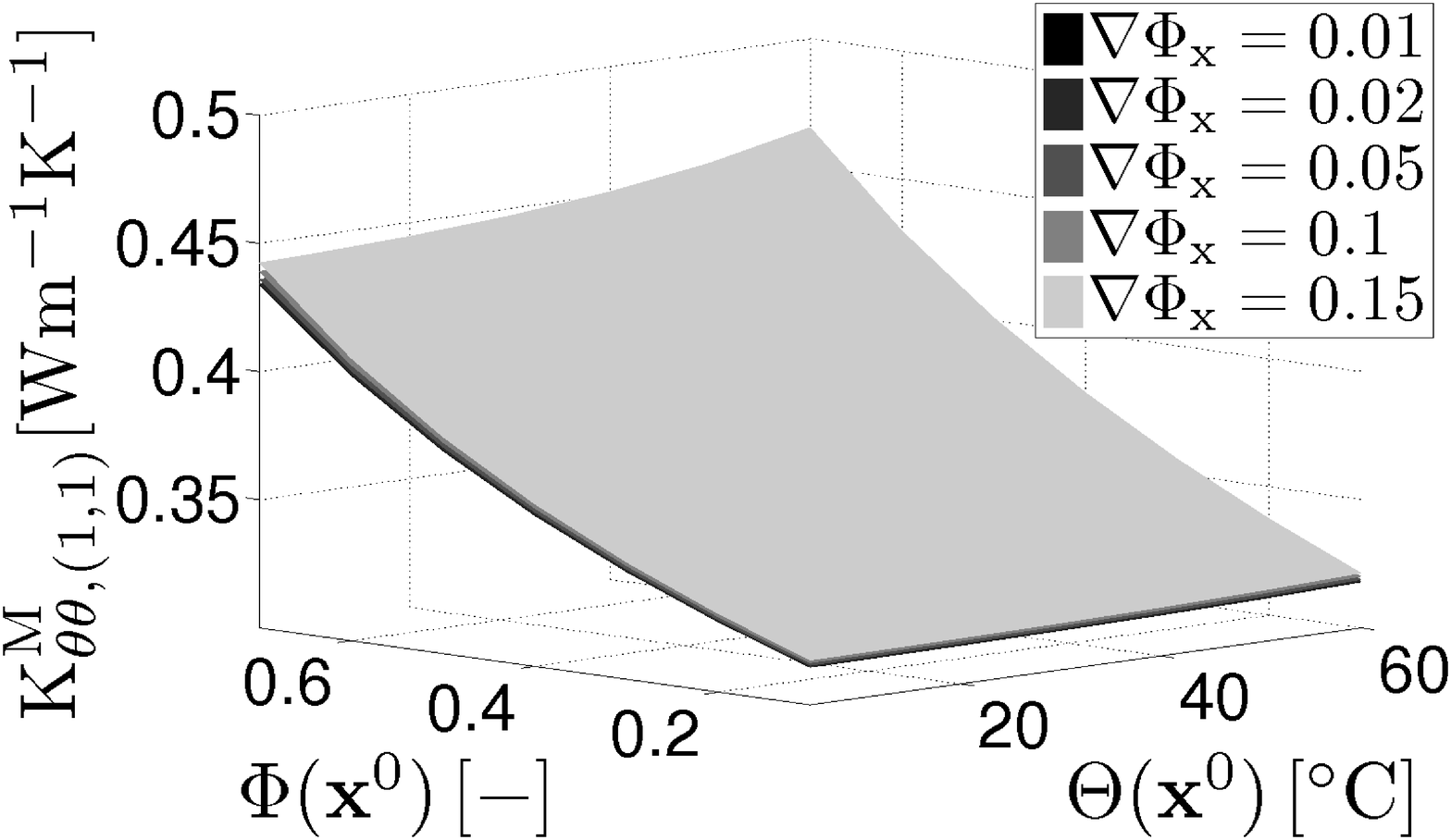}
& \includegraphics[width=57mm,keepaspectratio]{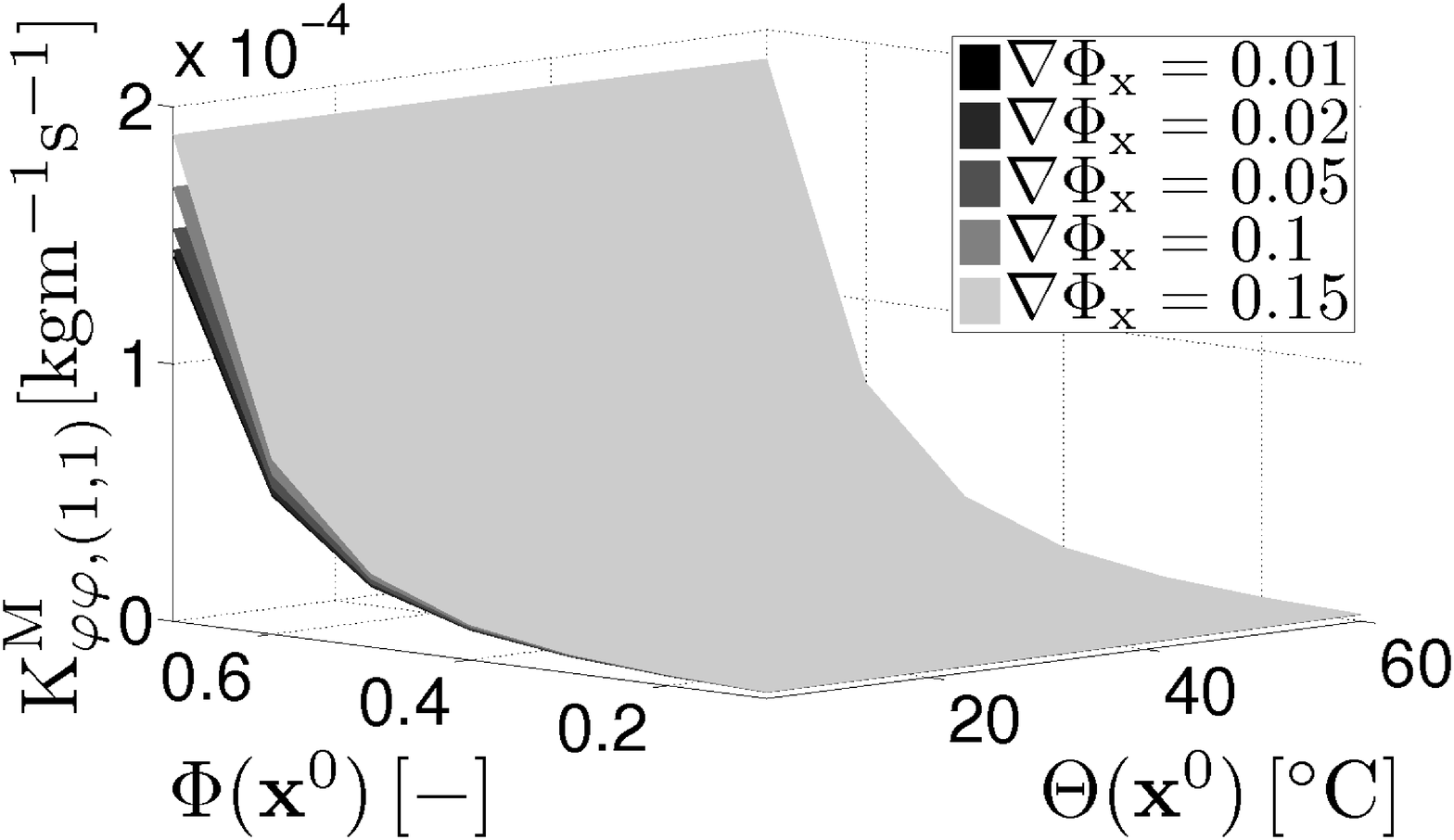}\\
(a) & (b)\\
\end{tabular}
\caption{Variation of the effective parameters
  ($\mathrm{K}_{\theta\theta,(1,1)}^{\mathrm{M}}$ and
  $\mathrm{K}_{\varphi\varphi,(1,1)}^{\mathrm{M}}$) as a function of
  initial conditions for various values of relative humidity
  gradients} \label{fig:Rgrad}
\end{center}
\end{figure}

The distributions of the same macroscopic terms as in the first
example appear in Fig.~\ref{fig:Rgrad}. It is evident that the
predicted effective parameters are considerably dependent on both the
initial and loading conditions. Despite it, one may suggest that
homogenization analysis performed for various values of water content
$w$ and certain referenced initial values of temperature and relative
humidity can be exploited to construct the homogenized macro-scale
retention curves providing the analysis is independent of the applied
macroscopic gradients. Such curves would be then used in an
independent macroscopic study. While this seems acceptable for
effective thermal conductivities, this approach evidently fails for
effective moisture transport coefficients, which strongly depend on
the current temperature and relative humidity gradients. Therefore,
studying the coupled heat and moisture transport in masonry structures
must be envisioned in a full-fledged coupled multi-scale framework
(FE$^2$ problem). On the other hand, this offers the possibility of
including the mesostructural morphology and mesostructural material
behavior in the macro-level, where typical structures are analyzed,
without the need for assigning the fine-scale details to the entire
structure.

\section{Conclusions}
\label{sec:conclusion}
A coupled heat and moisture transport was studied with reference to
the first order homogenization of masonry walls. First, a specific
experimental program was executed to infer the local phase material
parameters and interface transition coefficients from experimentally
observed jumps in temperature and relative humidity fields. The notion
of a negligible effect of considering an imperfect hydraulic contact
put forward already by experimental results was further supported
numerically by a parametric meso-scale homogenization study of a
stationary problem.

Admitting only a perfect hydraulic contact proves useful in the light
of the second set of results promoting the need for fully coupled
multi-scale analysis when transport of moisture becomes appreciable,
such as the case of historic masonry structures, especially bridges.

\section*{Acknowledgment}
This outcome has been achieved with the financial support of the
Ministry of Education, Youth and Sports, project No. 1M0579, within
activities of the CIDEAS research centre. In this undertaking,
theoretical results gained in the project 103/08/1531 were partially
exploited. The experimental work performed at the Department of
materials and chemistry of the Czech Technical University in Prague,
Faculty of Civil Engineering under the leadership of Prof. Robert
\v{C}ern\'{y} is also gratefully acknowledged.

\bibliographystyle{elsarticle-num}
\bibliography{liter-paper}
\end{document}